\documentclass[twocolumn]{aastex631}

\submitjournal{ApJ}
\usepackage{cuted}
\usepackage{amsmath}
\usepackage{booktabs}
\usepackage{float}
\usepackage{longtable}
\usepackage{placeins}

\begin{document}

\title{The $230$\,GHz Variability of Numerical Models of Sagittarius~A*  \\
I. Parameter Surveys on Varying Ion-electron Temperature Ratios Under Strongly Magnetized Conditions}

\author[0000-0003-2776-082X]{Ho-Sang Chan}
\altaffiliation{Croucher Scholar}
\affiliation{Department of Physics and Institute of Theoretical Physics, The Chinese University of Hong Kong, Shatin, N.T., Hong Kong}
\affiliation{JILA, University of Colorado and National Institute of Standards and Technology, 440 UCB, Boulder, CO 80309-0440, USA}
\affiliation{Department of Astrophysical and Planetary Sciences, University of Colorado, 391 UCB, Boulder, CO 80309, USA}

\author[0000-0001-6337-6126]{Chi-kwan Chan}
\affiliation{Steward Observatory and Department of Astronomy, University of Arizona, 933 N. Cherry Avenue, Tucson, AZ 85721, USA}
\affiliation{Data Science Institute, University of Arizona, 1230 N. Cherry Avenue, Tucson, AZ 85721, USA}
\affiliation{Program in Applied Mathematics, University of Arizona, 617 North Santa Rita, Tucson, AZ 85721, USA}

\author[0000-0002-0393-7734]{Ben S. Prather}
\affiliation{Los Alamos National Lab, Los Alamos, NM, 87545, USA}

\author[0000-0001-6952-2147]{George N. Wong}
\affiliation{School of Natural Sciences, Institute for Advanced Study, 1 Einstein Drive, Princeton, NJ 08540, USA}
\affiliation{Princeton Gravity Initiative, Princeton University, Princeton, NJ 08544, USA}

\author[0000-0001-7451-8935]{Charles Gammie}
\affiliation{Department of Physics, University of Illinois at Urbana-Champaign, 1110 West Green Street, Urbana, IL 61801, USA}
\affiliation{Department of Astronomy, University of Illinois at Urbana-Champaign, 1002 West Green Street, Urbana, IL 61801, USA}
\affiliation{NCSA, University of Illinois at Urbana-Champaign, 1205 W. Clark St., Urbana, IL 61801, USA}
\affiliation{Illinois Center for the Advanced Study of the Universe, University of Illinois at Urbana-Champaign, 1110 West Green St., Urbana, IL 61801, USA}

%%%%%%%%%%%%%%%%%%%%%%%%%%%%%%%%%%%%%%%%%%%%%%%%%%%%%%%%%%%%%%%%%%%%%%%%%%%%%%%%%%%%%%%%%%%%%%%%%%%%%%%%%%%%%%%%%%%%%

\begin{abstract}

The $230$\,GHz lightcurves of Sagittarius~A* (Sgr~A*) predicted by general relativistic magnetohydrodynamics (GRMHD) and ray-tracing (GRRT) models in \citet{2022ApJ...930L..16E} have higher variability $M_{\Delta T}$ compared to observations. In this series of papers, we explore the origin of such large brightness variability. In this first paper, we performed large GRRT parameter surveys that span from the optically thin to the optically thick regimes, covering the ion-electron temperature ratio under strongly magnetized conditions, $R_{\rm Low}$, from $1$ to $60$. We find that increasing $R_{\rm Low}$ can lead to either an increase or a reduction in $M_{\Delta T}$ depending on other model parameters, making it consistent with the observed variability of Sgr~A* in some cases. Our analysis of GRRT image snapshots finds that the major contribution to the large $M_{\Delta T}$ for the $R_{\rm Low} = 1$ models comes from the photon rings. However, secondary contributions from the accretion flow are also visible depending on the spin parameter. Our work demonstrates the importance of the electron temperature used for modelling radiatively inefficient accretion flows and places new constraints on the ion-electron temperature ratio. A more in-depth analysis for understanding the dependencies of $M_{\Delta T}$ on $R_{\rm Low}$ will be performed in subsequent papers.

\end{abstract}


%%%%%%%%%%%%%%%%%%%%%%%%%%%%%%%%%%%%%%%%%%%%%%%%%%%%%%%%%%%%%%%%%%%%%%%%%%%%%%%%%%%%%%%%%%%%%%%%%%%%%%%%%%%%%%%%%%%%%

\section{Introduction} \label{sec:intro}

The nature of the centre of our Galaxy, Sagittarius~A* (Sgr~A*), was a mystery for decades. Although radio \citep{lo1993high} and X-ray observations \citep{baganoff2003chandra}, as well as stellar proper motion data \citep{eckart2002stellar}, suggested that it could be a supermassive black hole, direct evidence was lacking. It was not until the recent progress made by the Event Horizon Telescope (EHT) Collaboration \citep{akiyama2022first} that the mystery of Sgr~A* was unveiled. By directly imaging the black hole accretion disk, it is now confirmed that Sgr~A* is a low-luminosity accreting supermassive black hole.

Sgr~A* is estimated to be accreting at a sub-Eddington rate of $\sim 10^{-5}\,M_{\odot} \mathrm{year}^{-1}$ \citep{quataert1999accretion, quataert2000constraining}. The low luminosity of Sgr~A* indicates that it is a radiatively inefficient accretion flow (RIAF) --- the accretion disc is believed to be geometrically thick and optically thin while transporting energy to the black hole mainly through advection. Under such circumstances, electrons are decoupled from the ions, and the plasma attains a two-temperature state \citep{1976ApJ...204..187S}, because \emph{i}) the electrons are unable to cool efficiently by radiation and \emph{ii}) Coulomb coupling between the ions and the electrons is weak \citep{begelman2014accreting}. In addition, electrons could attain a non-thermal \citep{2000ApJ...541..234O} number density distribution presumably through acceleration by shocks \citep{2011ApJ...726...75S}, magnetic reconnections \citep{werner2018non}, and/or turbulence \citep{zhdankin2019electron}, and electrons are unable to be thermalized easily due to weak Coulomb coupling and low cooling efficiency. When modelling the emission from RIAF, accurate electron distributions and precise temperatures are required. However, both are still poorly understood\footnote{Note that some recent progress has been made in better understanding the electron distributions in RIAF. See, for instance, \citet{scepi2022sgr}}. To the first-order approximation, it is reasonable to parameterize an energy partition between the electrons and ions. We assume the ion temperature $T_{i}$ is a fraction $R$ of the electron temperature $T_{e}$. The fraction $R$ could depend on the local environment and/or global accretion flow properties. 

Earlier work, such as that by \citet{2009ApJ...703L.142D}, assumed $R$ was constant over the entire domain, producing reasonable agreement with the millimetre VLBI visibility of Sgr~A*. However, it was soon found to overproduce near-infrared and X-ray fluxes \citep{2009ApJ...706..497M}. Following this, \citet{chan2015power} proposed a prescription function where $R$ depends on the plasma magnetization $\beta = 2P/|B|^{2}$, where $P$ is the gas pressure and $B$ is the magnetic field strength. In particular, $R$ is a step-function of $\beta$ at some threshold $\beta_{\rm Crit}$. This step-function prescription helps clearly distinguish emissions coming from the disk and the jet, where $\beta$ differs drastically. It was motivated by the fact that the electron temperature can drastically differ from the ion's when the density is low, and the magnetic field is strong. Using this prescription function, \citet{chan2015power} obtained parameter spaces where suites of accreting black hole models agree with observations. 

The electron and ion temperature partitioning used by \citet{2022ApJ...930L..16E} in their theoretical studies of Sgr~A* reads:

\begin{equation} \label{eqn:fraction}
    R = \frac{T_{i}}{T_{e}} = \frac{R_{\rm High}b^{2} + R_{\rm Low}}{b^{2} + 1},
\end{equation}

where $b = \beta/\beta_{\rm Crit}$. This prescription was first used by \citet{moscibrodzka2016general}, where $R_{\rm High}$ ($R_{\rm Low}$) represent the ion-electron temperature ratios in the weakly (strongly) magnetized regime. This function smooths the sharp transition of the step-function prescription by \citet{chan2015power}. It was motivated by the results of particle-in-cell simulations, which suggest that collisionless plasma preferentially heat the ions for $\beta > 1$ \citep[see][and references therein]{akiyama2019first}. The power-law dependencies on $b$ guarantee that radiations from strong and weak ion-electron coupling regions are easily distinguishable. 

In \citet{2022ApJ...930L..16E}, both $\beta_{\rm Crit}$ and $R_{\rm Low}$ are fixed at a constant value of $1$. This simple description yielded results for accreting Sgr~A* images that passed almost all constraints, such as the image size, visibility amplitude morphology, M-rings fits, infrared and near-infrared fluxes, etc., except for the $230$\,GHz flux variability. The variability is quantified as the ratio $M_{\Delta T} = \sigma_{\Delta T}/\mu_{\Delta T}$, where $\sigma_{\Delta T}$ is the standard deviation and $\mu_{\Delta T}$ is the mean of the $230$\,GHz flux over a time $\Delta T$. For the EHT Collaboration study, $\Delta T = 530\,GMc^{-2} = 3\,\mathrm{hours}$. The constraint of $M_{\Delta T}$ for Sgr~A* could not be met by most of the fiducial models in \citet{2022ApJ...930L..16E}. The reason why the $230$\,GHz flux of the \textit{theoretical models} obtained through general-relativistic magnetohydrodynamics (GRMHD) simulations and general-relativistic ray-tracing (GRRT) post-processing are more variable than observations remains a mystery. 

\begin{figure}[t]
    \centering
    \includegraphics[width=1.0\linewidth]{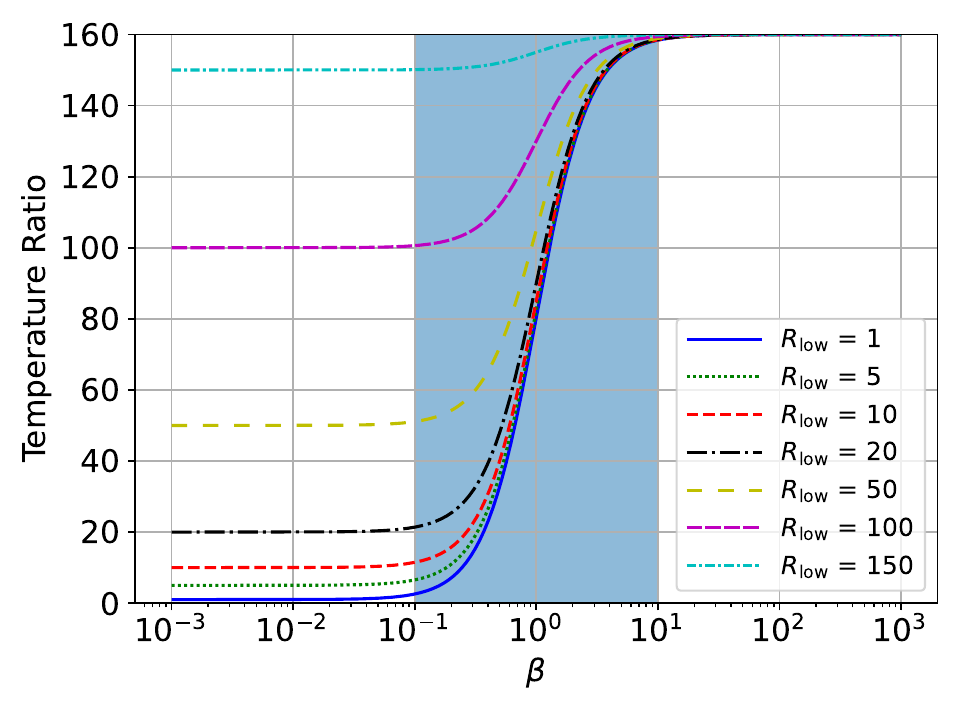}
    \caption{Visualization of Equation~(\ref{eqn:fraction}) with increasing $R_{\rm Low}$. In this case, $R_{\rm High}$ is fixed at $160$, and we assume $\beta_{\rm Crit} = 1$. The blue-shaded region indicates where $R$ is sensitive to variations in $\beta$. \label{fig:rhighrlow}}
\end{figure}

Here, we investigate whether varying $R_{\rm Low}$ can reduce the variability of the $230$\,GHz flux of the theoretical models of Sgr~A*. The motivation for varying $R_{\rm Low}$ can be understood in two ways. First, it is the only parameter that has not been studied in-depth before, and it is natural to investigate, with a wider parameter space, if it has any effect on $M_{\Delta T}$. Second, we can consider the functional form of Equation~\ref{eqn:fraction}, which is visualized in Figure~\ref{fig:rhighrlow}. We find that $R$ undergoes a sharp transition around $\beta = 10^{-1}$ and $10^{1}$, for $\beta_{\rm Crit} = 1$ and $R_{\rm High} = 160$. Given that the major contributions to the electromagnetic emissions at $230$\,GHz are mostly plasma located near the horizon \citep{dexter2020parameter}, where $\beta$ has a similar order of magnitude, a slight fluctuation in $\beta$ could lead to a drastic change in $R$, and thus $T_{e}$, potentially inducing high variability in the electromagnetic emission. By varying $R_{\rm Low}$, one can narrow the gap between $R_{\rm High}$ and $R_{\rm Low}$, making $R$ less sensitive to $\beta$ around $10^{-1}$ and $10^{1}$, and potentially solving the variability problem. We note that \citet{2022ApJ...930L..16E} increased $R_{\rm Low}$ up to $10$ and found no systematic reduction in the $230$\,GHz variability. Thus, to extend their work, we will further increase $R_{\rm Low}$ to explore a wider range of parameter spaces. This paper is structured as follows: In Section~\ref{sec:method}, we describe our methods, namely the GRMHD simulation libraries of Sgr~A*, the tools for GRRT parameter surveys, and the parameter spaces that we consider. In Section~\ref{sec:results}, we present the results of our parameter searches, showing how $M_{\Delta T}$ changes as $R_{\rm Low}$ is increased for theoretical models of Sgr~A*. In fact, we find a \textit{reduction} of $M_{\Delta T}$, and we will compare the reduced value with observations. In the same section, we will also unravel why GRMHD models of Sgr~A* are more variable than observations. Finally, we conclude our studies in Section~\ref{sec:conclu}.

%%%%%%%%%%%%%%%%%%%%%%%%%%%%%%%%%%%%%%%%%%%%%%%%%%%%%%%%%%%%%%%%%%%%%%%%%%%%%%%%%%%%%%%%%%%%%%%%%%%%%%%%%%%%%%%%%%%%%

\section{Methodology} \label{sec:method}

This section describes our methods of obtaining the GRMHD models, performing the GRRT parameter surveys, and constructing the GRRT image libraries of Sgr~A* with varying $R_{\rm Low}$. 

%%%%%%%%%%%%%%%%%%%%%%%%%%%%%%%%%%%%%%%%%%%%%%%%%%%%%%%%%%%%%%%%%%%%%%%%%%%%%%%%%%%%%%%%%%%%%%%%%%%%%%%%%%%%%%%%%%%%%

\subsection{GRMHD Simulations} \label{subsec:grmhd}

Our non-radiative GRMHD models of Sagittarius A* are obtained using the GRMHD code \textsc{kharma}, which is the GPU-enabled version of  \texttt{iharm3d} \citep{prather2021iharm3d}. \textsc{kharma} solves the ideal GRMHD equations (with $c = 1$) and the divergence-less constraint of magnetic fields using the flux-constrained transport scheme \citep{toth2000b}. \textsc{kharma} assumes the `spherical-polar' version of the Kerr-Schild coordinates with the metric tensor given as:

\vspace{-0.1em}
\begin{widetext}
\begin{equation}
g_{\mu\nu} = \begin{pmatrix}
    -1 + 2r/\Sigma & 2r/\Sigma & 0 & -2ar\text{sin}^{2}\theta/\Sigma \\
    2r/\Sigma & 1 + 2r/\Sigma & 0 & -a\text{sin}^{2}\theta(1 + 2r/\Sigma) \\
    0 & 0 & \Sigma & 0 \\
    -2ar\text{sin}^{2}\theta/\Sigma & -a\text{sin}^{2}\theta(1 + 2r/\Sigma) & 0 & \text{sin}^{2}\theta(\Sigma + a^{2}\text{sin}^{2}\theta[1 + 2r/\Sigma])
\end{pmatrix}
\end{equation}
\end{widetext}

\noindent where $r$ and $\theta$ are spherical distances and polar angles. $\Sigma = r^{2} + a^{2}\text{cos}^{2}\theta$, where $a$ is the black-hole spin. \textsc{kharma} uses the WENO reconstruction method \citep{jiang1996efficient} to reconstruct primitive variables and the Lax-Friedrichs solver to solve the Riemann problem on cell boundaries. \textsc{kharma} evolves the GRMHD equations using the 2nd-order Strong Stability Preserving Runge-Kutta Method \citep{gottlieb2011strong} with a CFL number of $0.7$. We refer readers to \citet{wong2022patoka} to look for the detailed GRMHD equations being solved and their technical details.

The \textsc{kharma} simulation libraries for the EHT campaign include both the Standard and Normal Evolution (SANE) and Magnetic Arrested Disc (MAD) initial conditions. In this work, we consider only the MAD state since the MAD models are favoured in \citet{2022ApJ...930L..16E}. The accretion disc assumed an ideal gas equation of state with an adiabatic index of $4/3$, with the initial conditions being a geometrically thick torus computed using the method by \citet{1976ApJ...207..962F}. The initial torus has an inner radius at $20$ and a pressure maximum at $41$, all in units of $GMc^{-2}$. To initialize the MAD state, we set the initial magnetic field within the torus using the vector potential \citep{wong2022patoka}:

\begin{equation}
    A_{\phi} = \text{max} \left[\frac{\rho}{\rho_{\rm max}}\left(\frac{r}{r_{0}}\text{sin} \theta\right)e^{-r/400} - 0.2, 0 \right],
\end{equation} 

where $\rho_{\rm max}$ is the maximum density and $r_{0}$ is the inner radial boundary of the computational domain. Magnetic fields are computed using $\vec{B} = \nabla \times \vec{A}$. Finally, the simulation library employed the funky modified Kerr-Schild coordinates \citep{prather2022grmhd} to overcome time step restrictions along the pole. The radial outer boundary is located at 1,000$\,GMc^{-3}$, the simulation resolution is $288\times128\times128$, and the simulation duration is 30,000\,$GMc^{-2}$.

%%%%%%%%%%%%%%%%%%%%%%%%%%%%%%%%%%%%%%%%%%%%%%%%%%%%%%%%%%%%%%%%%%%%%%%%%%%%%%%%%%%%%%%%%%%%%%%%%%%%%%%%%%%%%%%%%%%%%

\subsection{GRRT} \label{subsec:grrt}

We compute the $230$\,GHz images and fluxes of the GRMHD models using the open-source GRRT code \texttt{IPOLE} \citep{moscibrodzka2018ipole}, which is based on the GRRT code \texttt{GRTRANS} \citep{dexter2016public}. In principle, \texttt{IPOLE} solves the polarized radiative transfer code, solving for the Stokes parameters ($I$, $Q$, $U$, $V$). However, the $Q$, $U$, and $V$ parameters are out of scope in this study. Thus, to reduce computational time cost, we solve only the unpolarized radiative transfer equation built-in \texttt{IPOLE}. The unpolarized, covariant radiative transfer equation reads:

\begin{equation} \label{eqn:transfer}
    \frac{d}{d\lambda}\left(\frac{I_{\nu}}{\nu^{3}}\right) = \frac{j_{\nu}}{\nu^{2}} - \alpha_{\nu}\frac{I_{\nu}}{\nu^{2}},
\end{equation}

where the subscript $\nu$ means the quantity for a given photon frequency, $I_{\nu}$ is the intensity, $j_{\nu}$ is the emissivity, and $\alpha_{\nu}$ is the absorptivity. $\lambda$ is the affine parameter along the photon geodesic:

\begin{align}
     \frac{dx^{\nu}}{d\lambda} &= k^{\nu}, \\
     \frac{dk^{\nu}}{d\lambda} &= -\Gamma^{\nu}_{\mu \sigma}k^{\mu}k^{\sigma}.
\end{align}

where $k^{\nu}$ is the photon four-vector. We use the transfer coefficients presented in \citet{dexter2016public} to solve Equation~\ref{eqn:transfer}. Following the work of \citet{2022ApJ...930L..16E}, we fix the black-hole image's field of view (FOV) to be $200$ muas. The resolution of the image is taken as $400 \times 400$ pixels, which is fine enough to ensure convergence. The distance from Earth to Sgr~A* is taken as 8,178\,parsec, while the mass of Sgr~A* is assumed to be $4.154 \times 10^{6}\,M_{\odot}$ \citep{abuter2019geometric}.

\begin{figure*}[htb]
    \centering
    \includegraphics[width=1.0\linewidth]{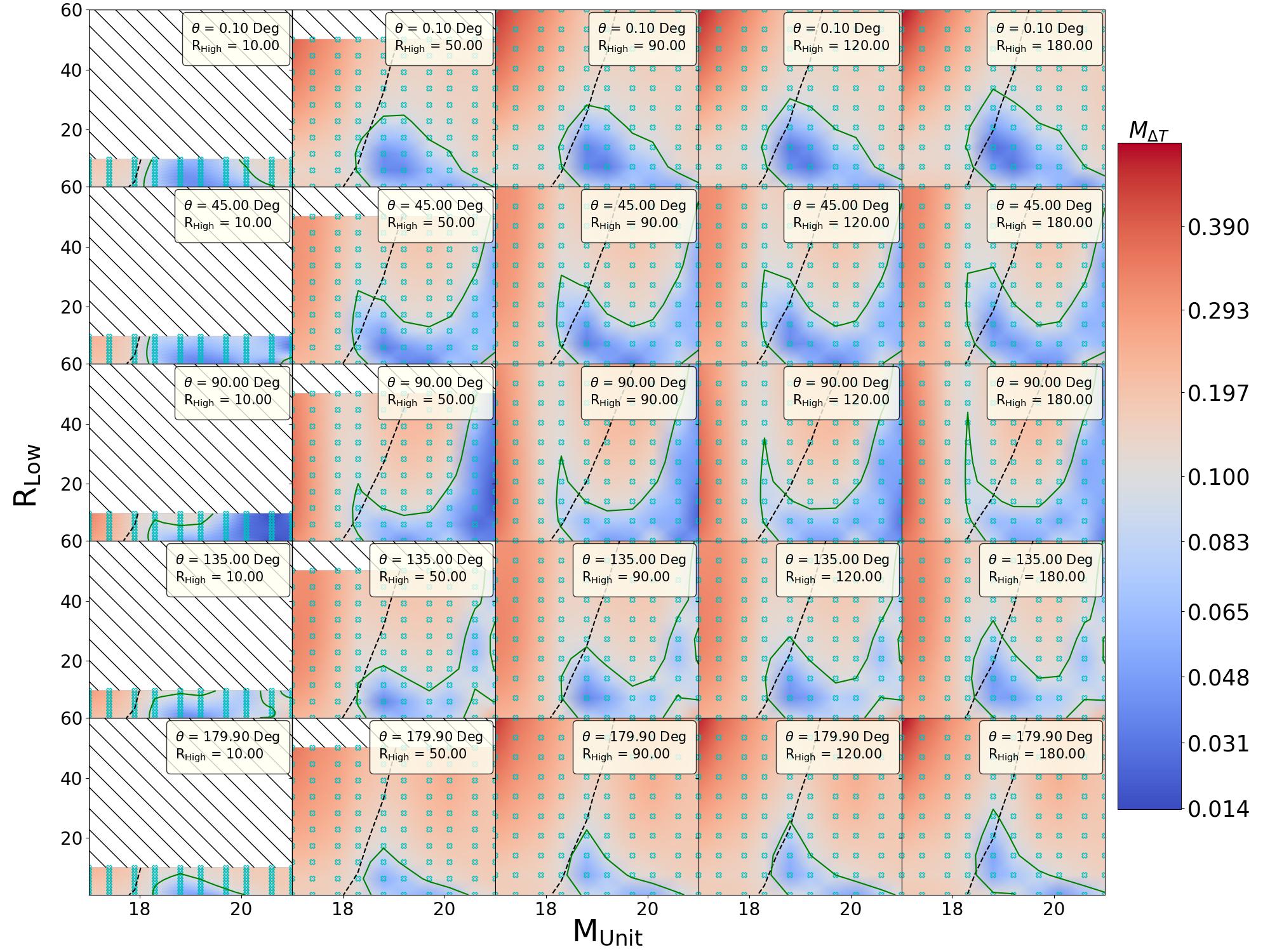}
    \caption{Example contour representations of $M_{\Delta T}$ against $R_{\rm Low}$ and $M_{\rm Unit}$ obtained through GRRT parameter surveys. This is for the GRMHD models with $a = +0.94$. The upper right box in each sub-grid shows the $\theta$ and $R_{\rm High}$ assumed for a particular set of GRRT. Also, the black-dotted contour lines represent models of Sgr~A* where the time-averaged flux is $2.4\,\textrm{Jy}$, which is the observed averaged fluxes of Sgr~A* at $230$\,GHz. For $R_{\rm High} = 10$ and $50$, the slashed region represent the parameter spaces where $R_{\rm High} < R_{\rm low}$, which is out of our interest. Light-blue markers indicate the point of models we obtained through GRRT. We use a diverging colormap that centers around $M_{\Delta T} = 0.1$, the upper limit of the historical distribution of $M_{\Delta T}$ for Sgr~A*. Thus, regions with red (blue) colours indicate models with high (low) $M_{\Delta T}$. We also overlay green contour lines to represent the boundary of the parameter space where $M_{\Delta T} \leq 0.1$. The contour plots for other spins are included in Figure Set 1. \label{fig:m3contour}}

    \figsetstart
    \figsetnum{1}
    \figsettitle{$M_{\Delta T}$ against $R_{\rm Low}$ and $M_{\rm Unit}$ for $a = +0.5, 0, -0.5,$ and $-0.94$.}
    
    \figsetgrpstart
    \figsetgrpnum{1.1}
    \figsetgrptitle{m3contoura+0.5}
    \figsetplot{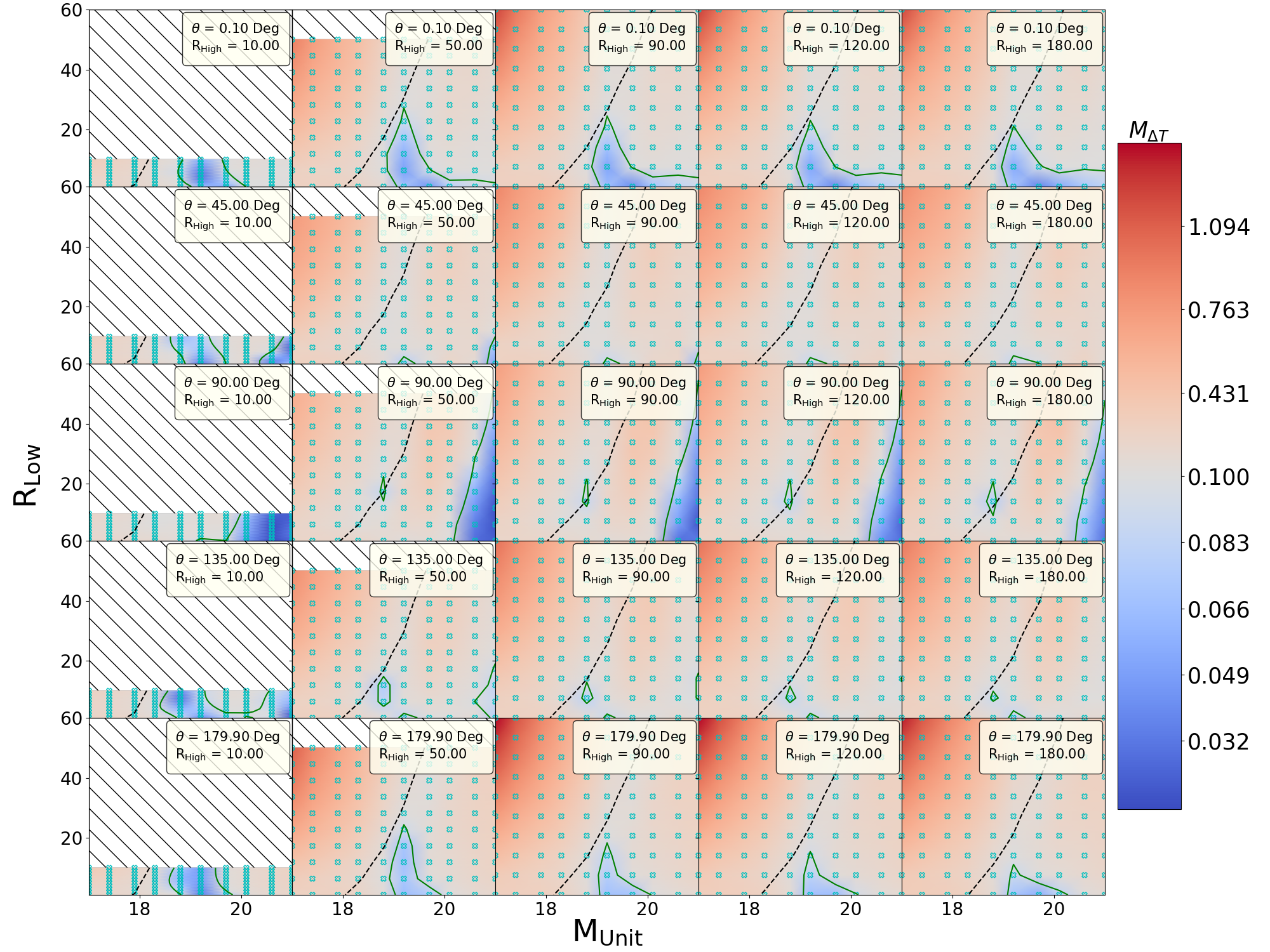}
    \figsetgrpnote{$M_{\Delta T}$ against $R_{\rm Low}$ and $M_{\rm Unit}$ for $a = +0.5$. See Figure~\ref{fig:m3contour} for the details.}
    \figsetgrpend

    \figsetgrpstart
    \figsetgrpnum{1.2}
    \figsetgrptitle{m3contoura0}
    \figsetplot{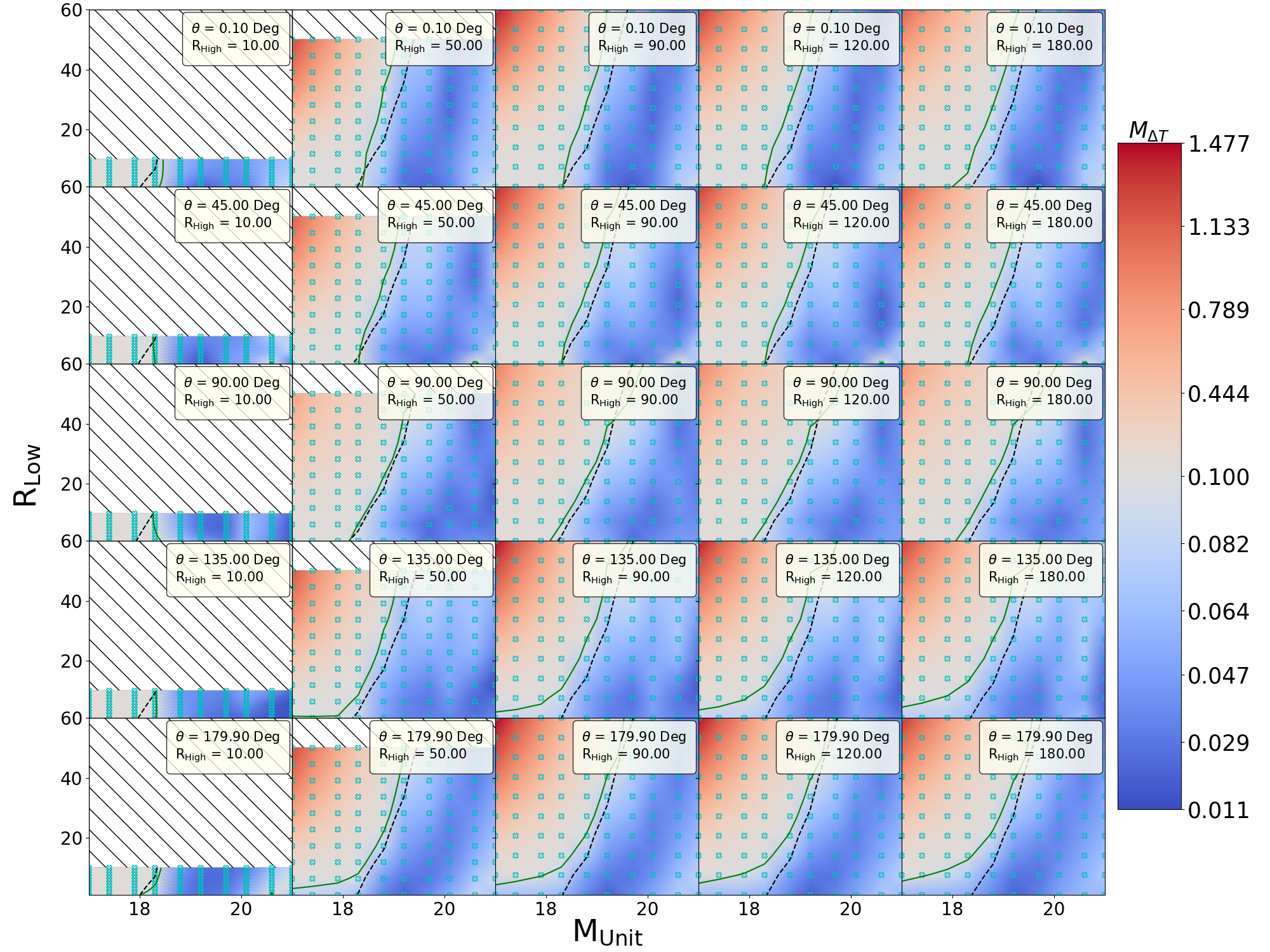}
    \figsetgrpnote{$M_{\Delta T}$ against $R_{\rm Low}$ and $M_{\rm Unit}$ for $a = 0$. See Figure~\ref{fig:m3contour} for the details.}
    \figsetgrpend

    \figsetgrpstart
    \figsetgrpnum{1.3}
    \figsetgrptitle{m3contoura-0.5}
    \figsetplot{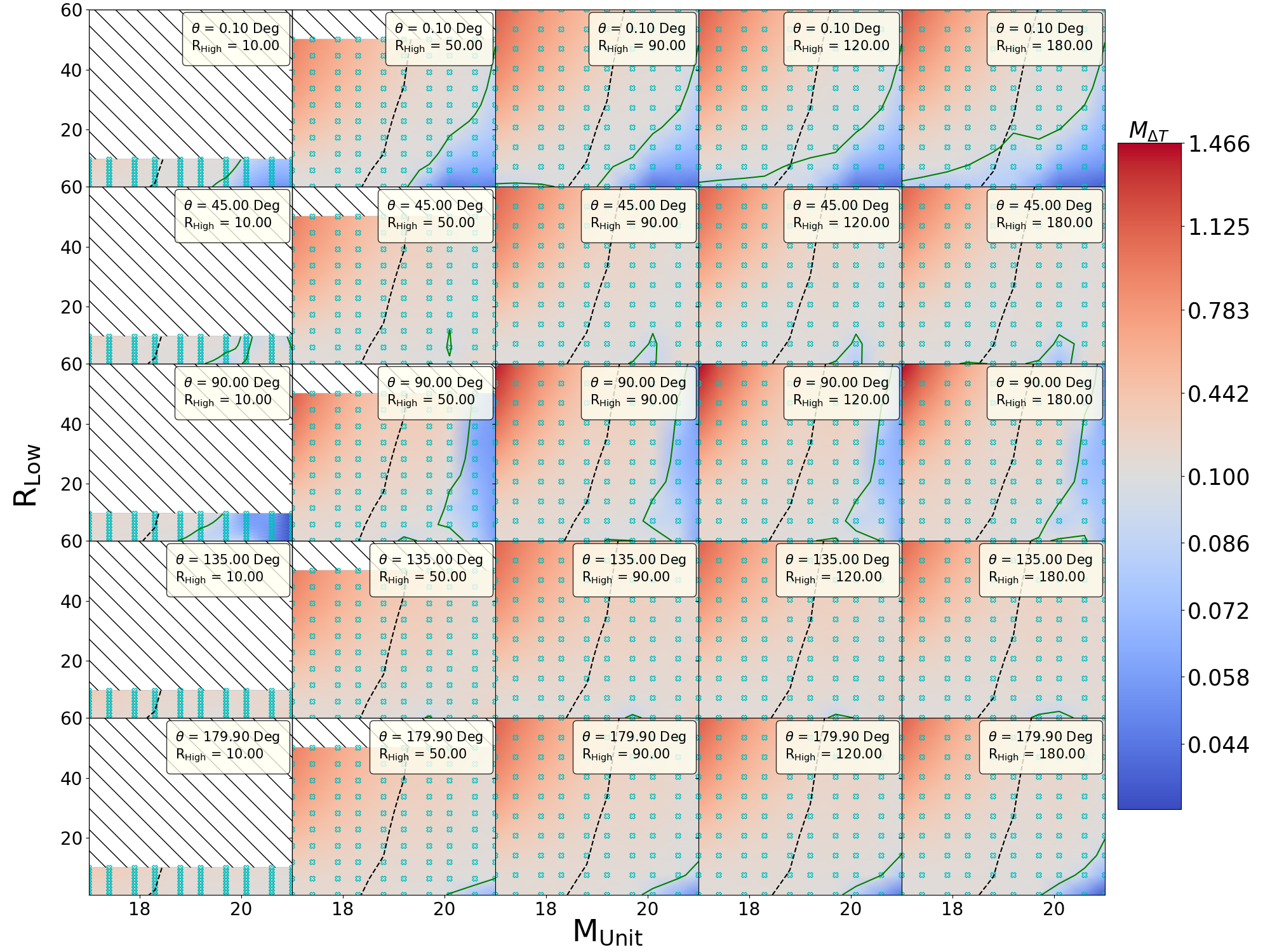}
    \figsetgrpnote{$M_{\Delta T}$ against $R_{\rm Low}$ and $M_{\rm Unit}$ for $a = -0.5$. See Figure~\ref{fig:m3contour} for the details.}
    \figsetgrpend

    \figsetgrpstart
    \figsetgrpnum{1.4}
    \figsetgrptitle{m3contoura-0.94}
    \figsetplot{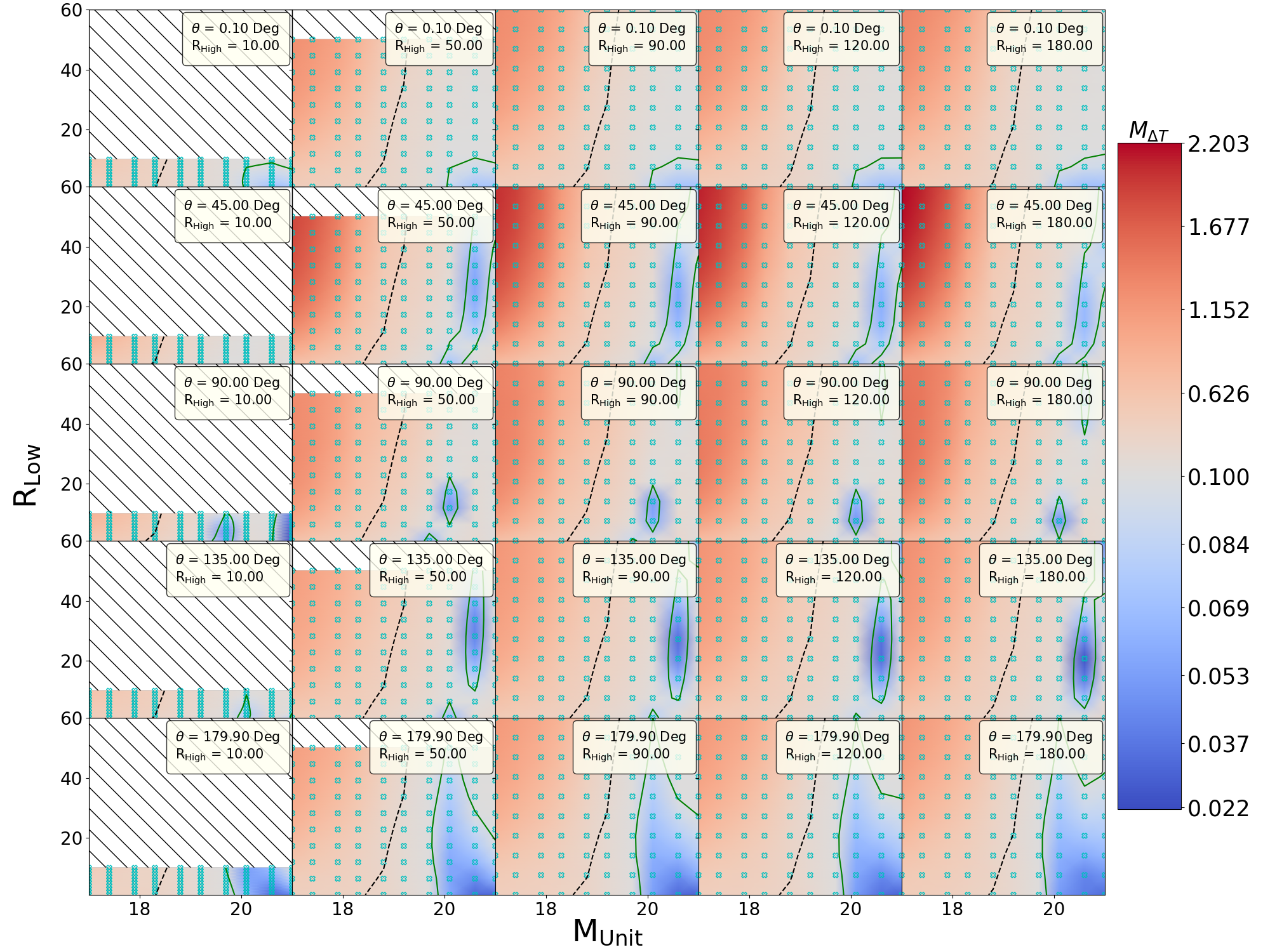}
    \figsetgrpnote{$M_{\Delta T}$ against $R_{\rm Low}$ and $M_{\rm Unit}$ for $a = -0.94$. See Figure~\ref{fig:m3contour} for the details.}
    \figsetgrpend
    
    \figsetend

\end{figure*}

\begin{figure*}[htb]
    \centering
    \gridline{\fig{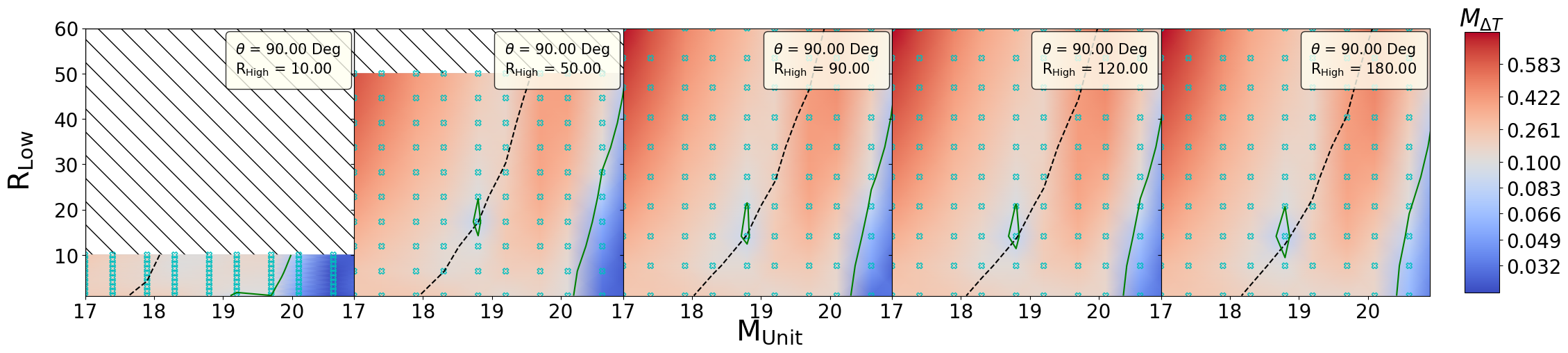}{0.9\textwidth}{(a) $a = +0.5, \theta = 90$ Deg}}
    \gridline{\fig{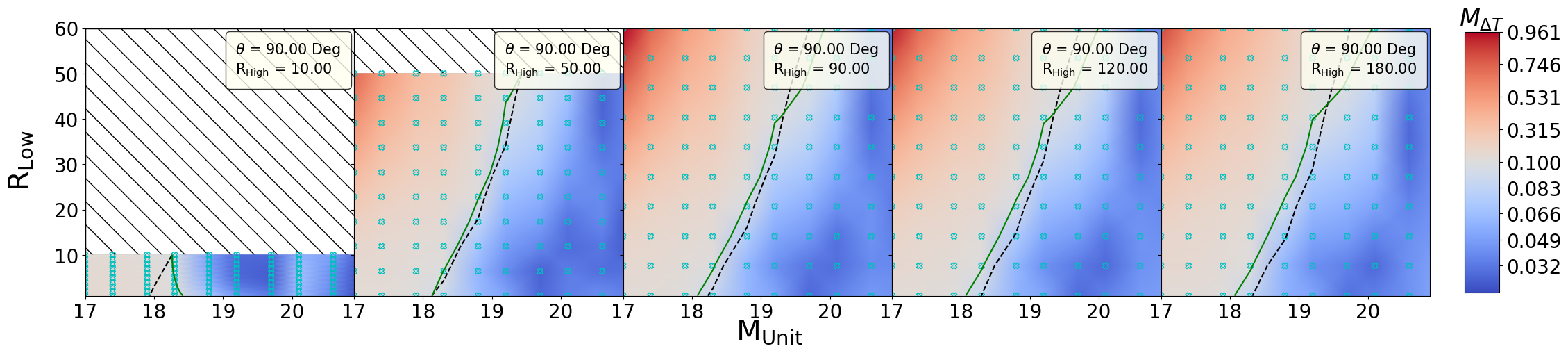}{0.9\textwidth}{(b) $a = 0, \theta = 90$ Deg}}
    \gridline{\fig{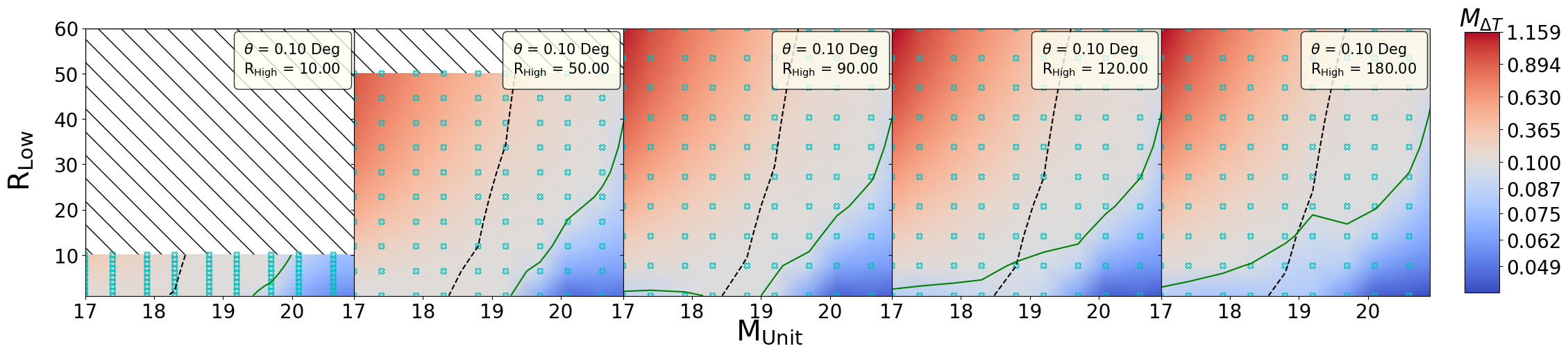}{0.9\textwidth}{(c) $a = -0.5, \theta = 0.1$ Deg}}
    \gridline{\fig{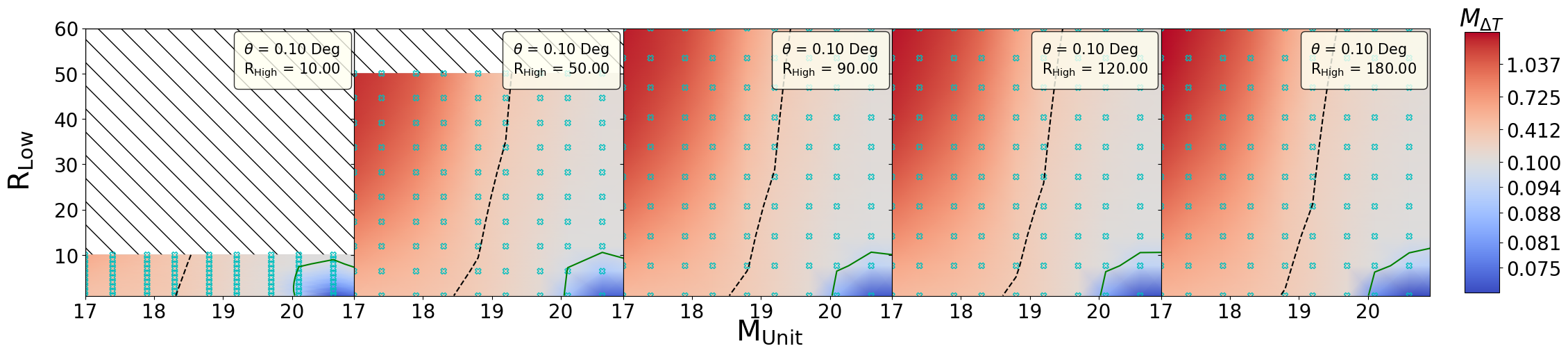}{0.9\textwidth}{(d) $a = -0.94, \theta = 0.1$ Deg}}
    \caption{Same as Figure~\ref{fig:m3contour}, but for the representative plots of each black hole spin $a$. We labelled $a$ and $\theta$ under the description of each subplot. \label{fig:m3contoursupp}}
\end{figure*}

\begin{figure*}[!htb]
    \centering
    \gridline{\fig{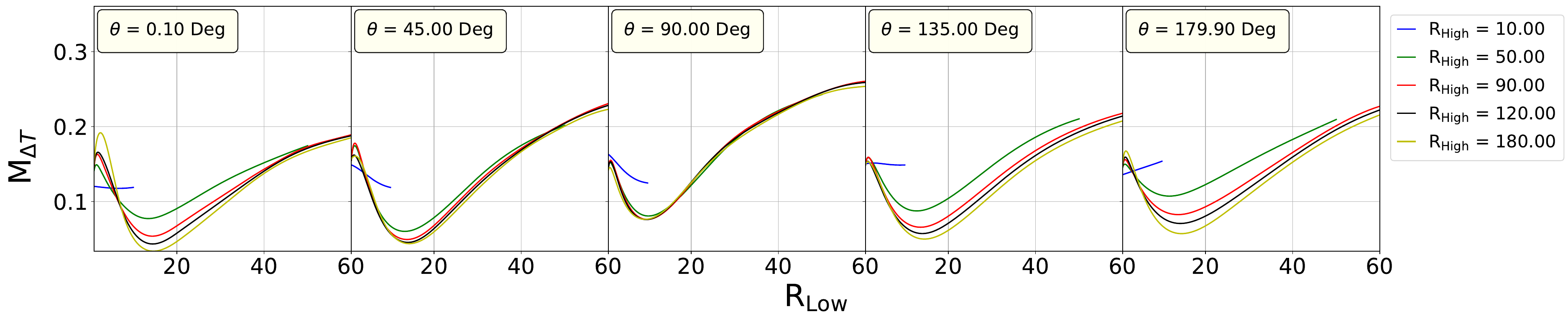}{0.9\textwidth}{(a) $a = +0.94$}}
    \gridline{\fig{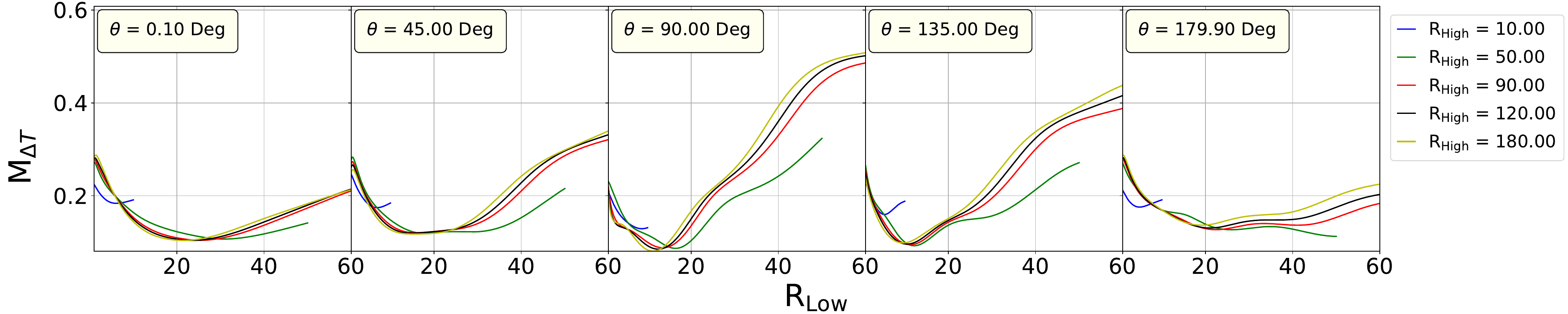}{0.9\textwidth}{(b) $a = +0.5$}}
    \gridline{\fig{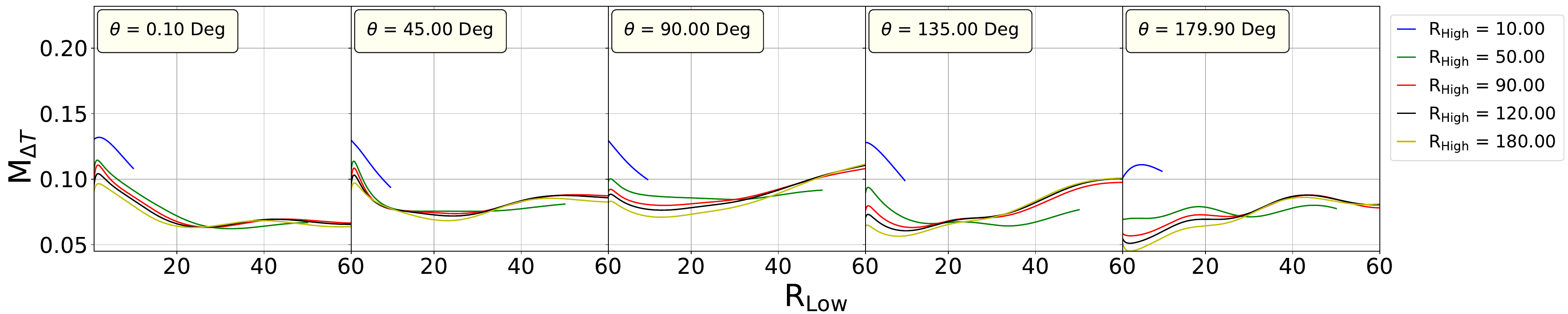}{0.9\textwidth}{(c) $a = 0$}}
    \gridline{\fig{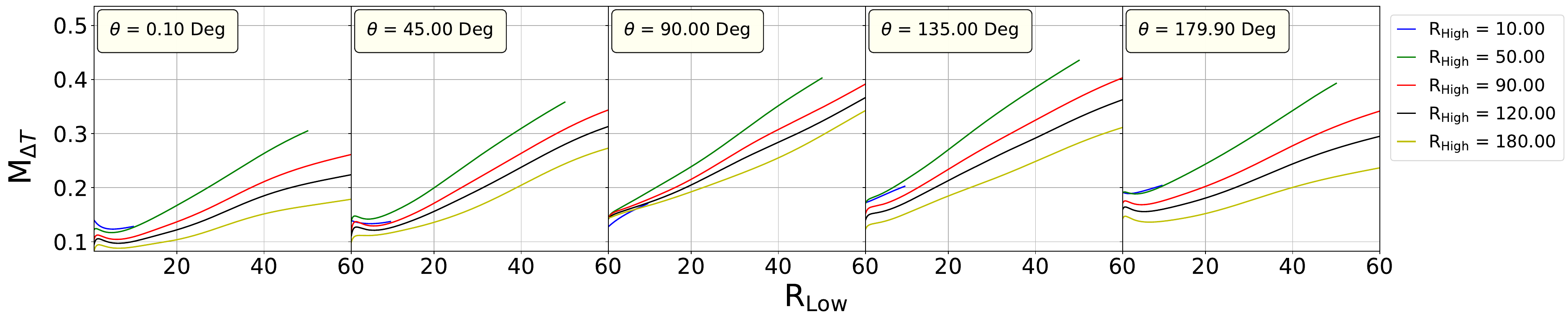}{0.9\textwidth}{(d) $a = -0.5$}}
    \gridline{\fig{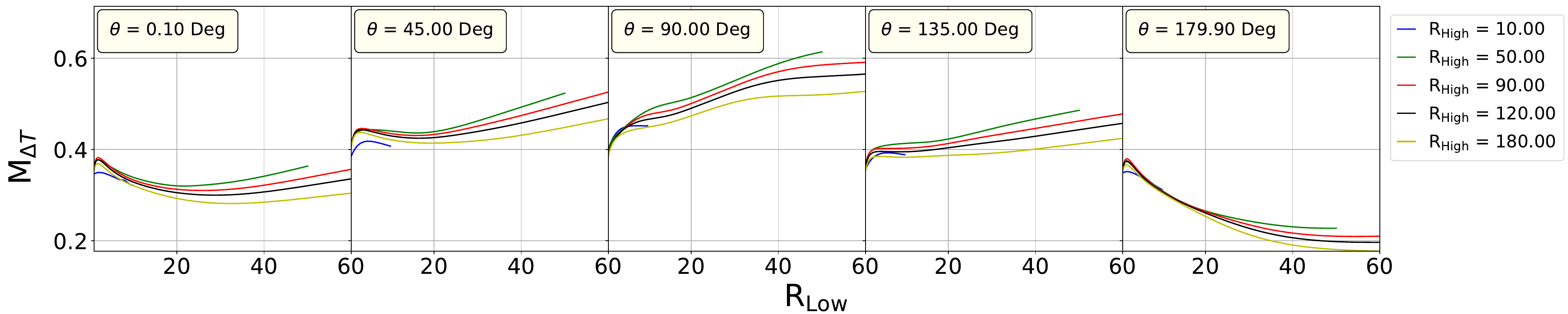}{0.9\textwidth}{(e) $a = -0.94$}}
    \caption{$M_{\Delta T}$ against $R_{\rm Low}$ for the black-hole models that satisfy the $2.4\,\textrm{Jy}$ constraint for all the black-hole spin we considered. Each sub-figure represents results for different $a$. The sub-grid plot in each sub-figure shows the variations of $M_{\Delta T}$ against $R_{\rm Low}$ for a given $\theta$ across different $R_{\rm High}$. Except for $R_{\rm High} = 10$, the shape of the $M_{\Delta T}$ versus $R_{\rm Low}$ curves are less sensitive across different $R_{\rm High}$ than $\theta$. The values of $M_{\Delta T}$ and $R_{\rm Low}$ are obtained from the parameter search results using Bivariate spline approximation provided by \texttt{Scipy} in the log-log scale. \label{fig:minm3}}
\end{figure*}

\begin{figure*}[htb]
    \centering
    \includegraphics[width=1.0\linewidth]{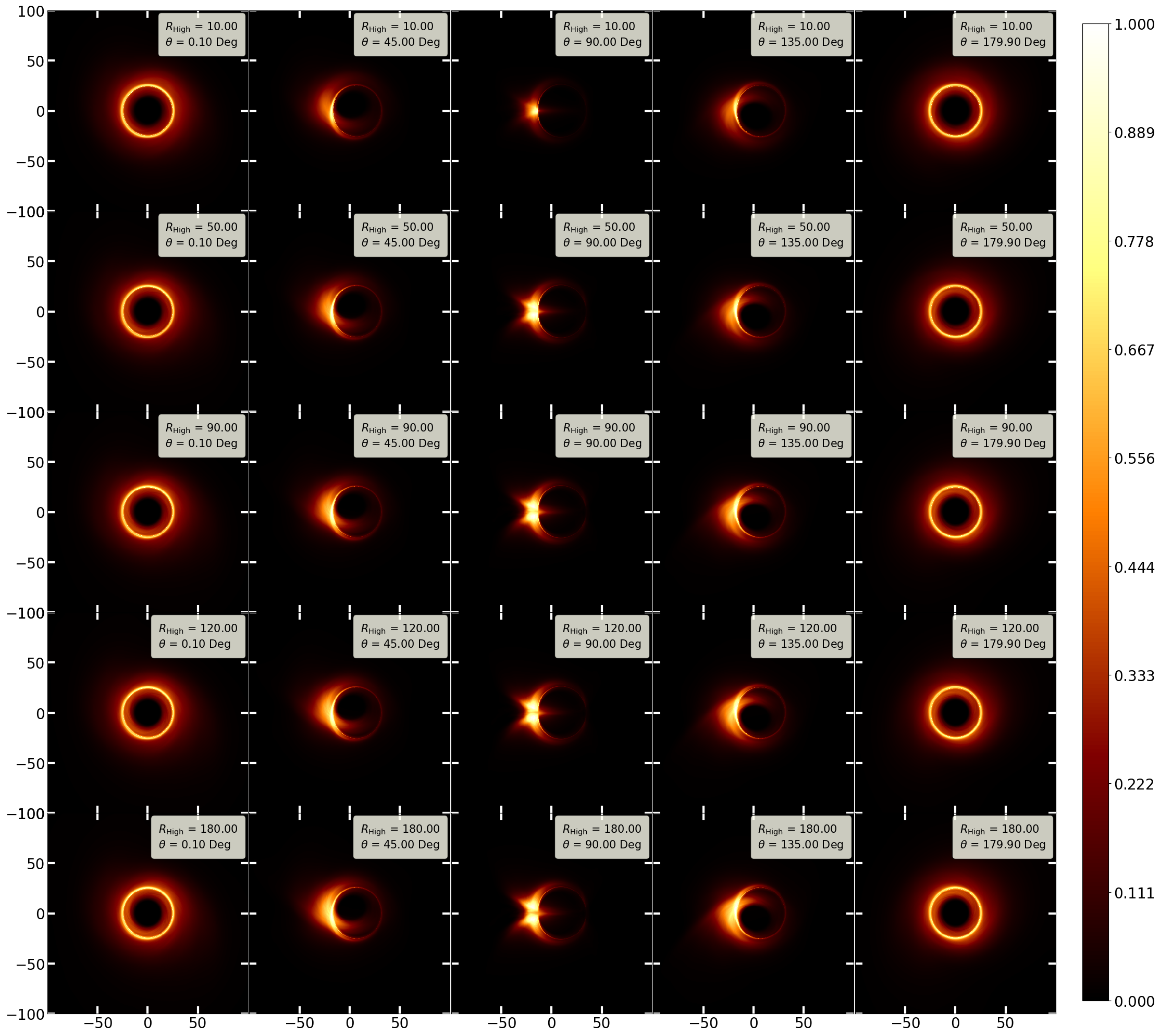}
    \caption{Example of time-averaged, pixel-wise luminosity plots. These images are computed with the GRMHD snapshots within the time interval of $\tau = (\textrm{29,465}$ -- $\textrm{29,995})\,GMc^{-2}$, while spanning different $R_{\rm High}$ and $\theta$. Here, $a = +0.94$, and we assume $R_{\rm Low} = 1$. The upper right box in each sub-grid shows the $\theta$ and $R_{\rm High}$ assumed. The colour map is scaled down to $[0,1]$ in each sub-plot. The luminosity plots for other spins are included in Figure Set 2. \label{fig:lumavg}}

    \figsetstart
    \figsetnum{2}
    \figsettitle{Time-averaged, pixel-wise luminosity across different $R_{\rm High}$ and $\theta$ (with $R_{\rm Low} = 1$) for $a = +0.5, 0, -0.5,$ and $-0.94$.}
    
    \figsetgrpstart
    \figsetgrpnum{2.1}
    \figsetgrptitle{lumavg+0.5}
    \figsetplot{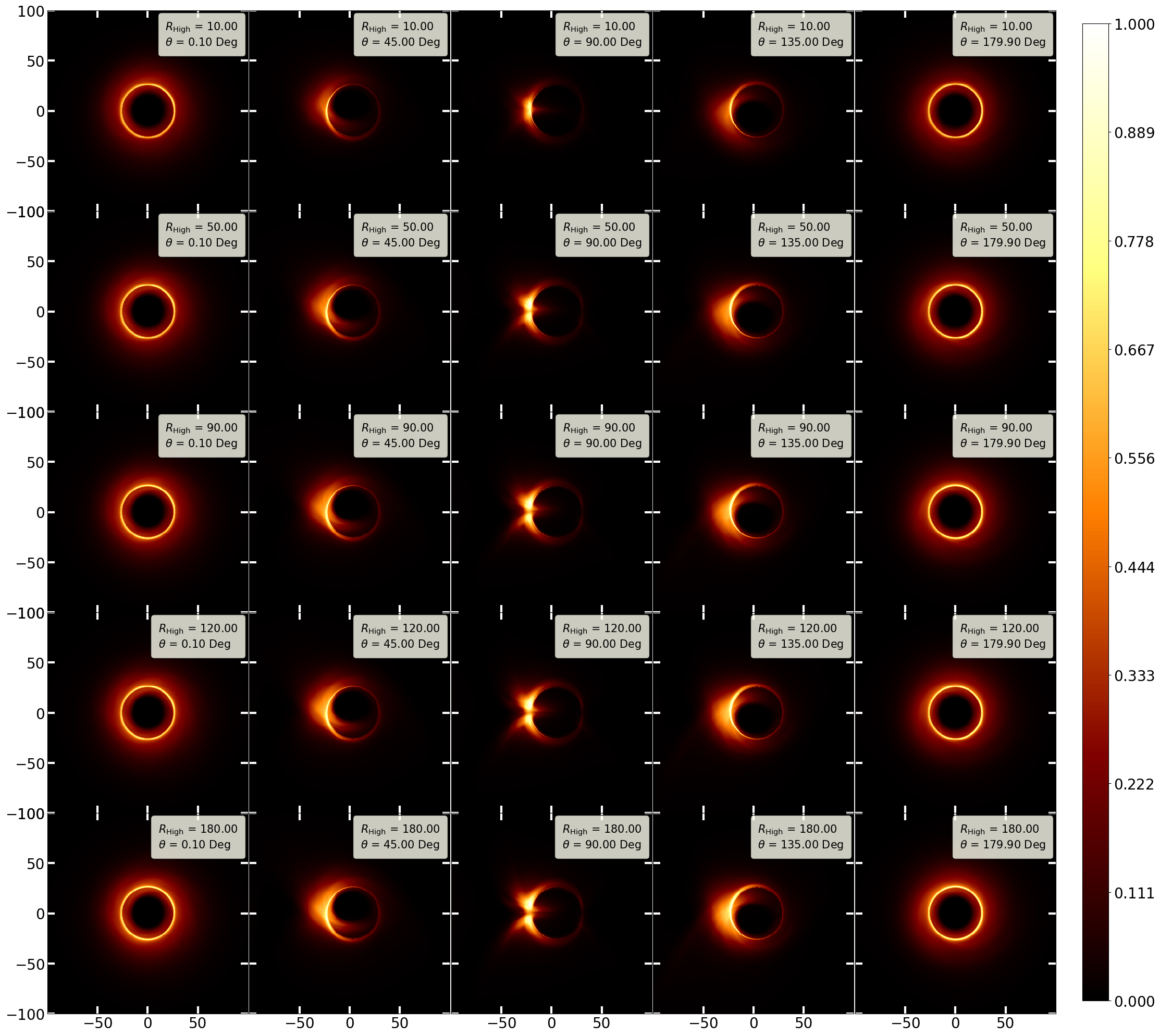}
    \figsetgrpnote{Time-averaged, pixel-wise luminosity across different $R_{\rm High}$ and $\theta$ (with $R_{\rm Low} = 1$) for $a = +0.5$. See Figure~\ref{fig:lumavg} for the details.}
    \figsetgrpend

    \figsetgrpstart
    \figsetgrpnum{2.2}
    \figsetgrptitle{lumavg0}
    \figsetplot{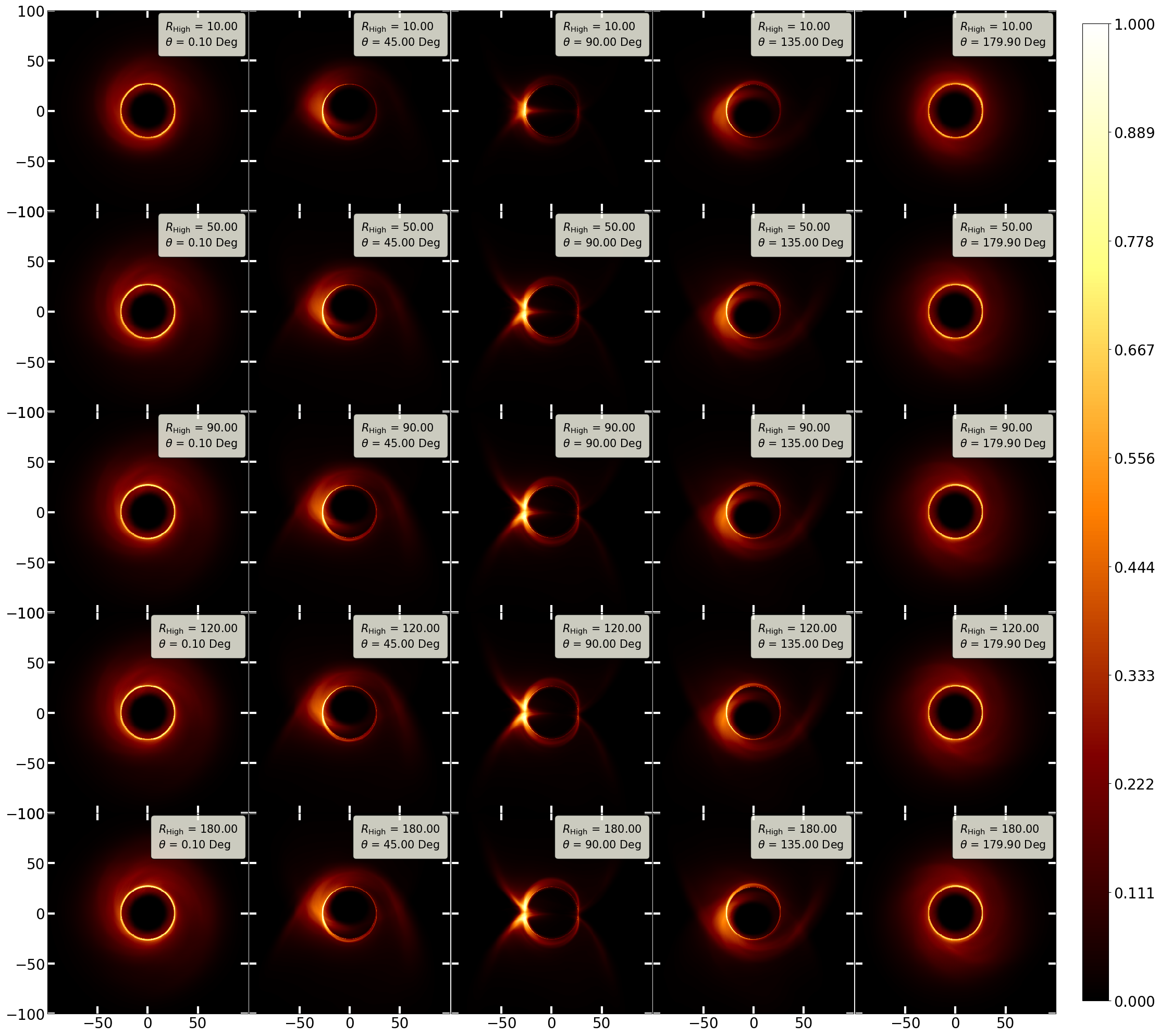}
    \figsetgrpnote{Time-averaged, pixel-wise luminosity across different $R_{\rm High}$ and $\theta$ (with $R_{\rm Low} = 1$) for $a = 0$. See Figure~\ref{fig:lumavg} for the details.}
    \figsetgrpend

    \figsetgrpstart
    \figsetgrpnum{2.3}
    \figsetgrptitle{lumavg-0.5}
    \figsetplot{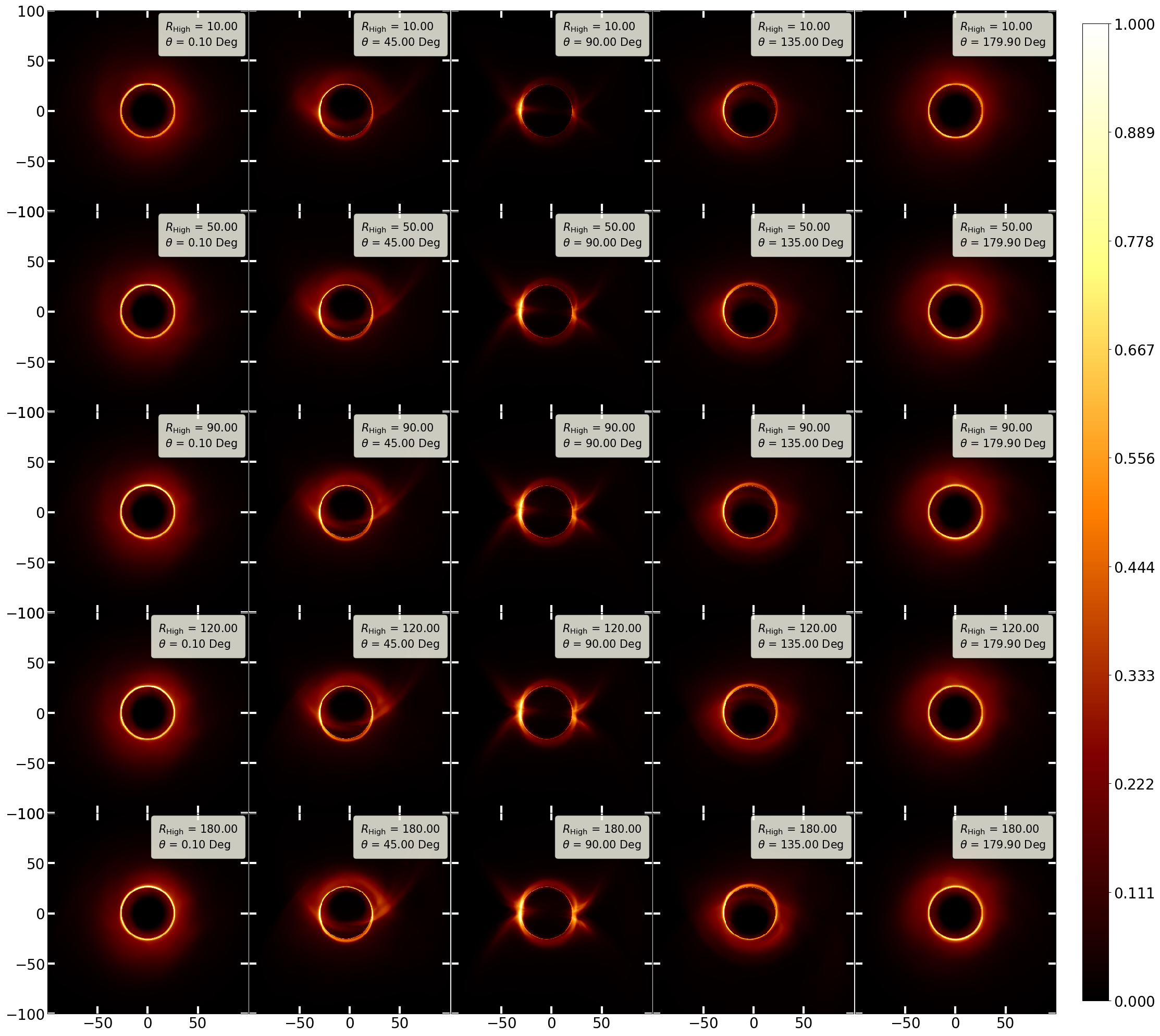}
    \figsetgrpnote{Time-averaged, pixel-wise luminosity across different $R_{\rm High}$ and $\theta$ (with $R_{\rm Low} = 1$) for $a = -0.5$. See Figure~\ref{fig:lumavg} for the details.}
    \figsetgrpend

    \figsetgrpstart
    \figsetgrpnum{2.4}
    \figsetgrptitle{lumavg-0.94}
    \figsetplot{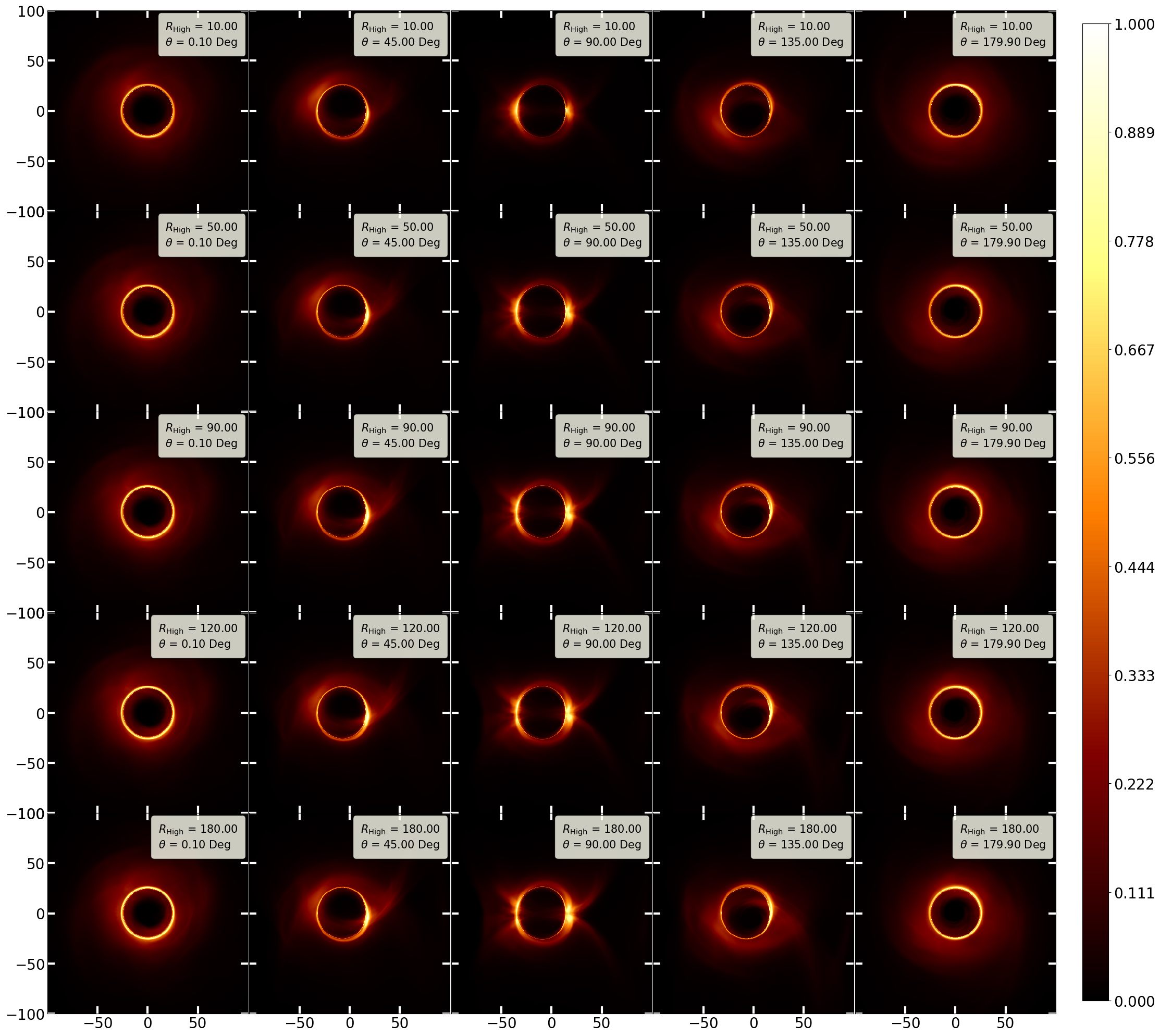}
    \figsetgrpnote{Time-averaged, pixel-wise luminosity across different $R_{\rm High}$ and $\theta$ (with $R_{\rm Low} = 1$) for $a = -0.94$. See Figure~\ref{fig:lumavg} for the details.}
    \figsetgrpend
    
    \figsetend

\end{figure*}

\begin{figure*}[htb]
    \centering
    \gridline{\fig{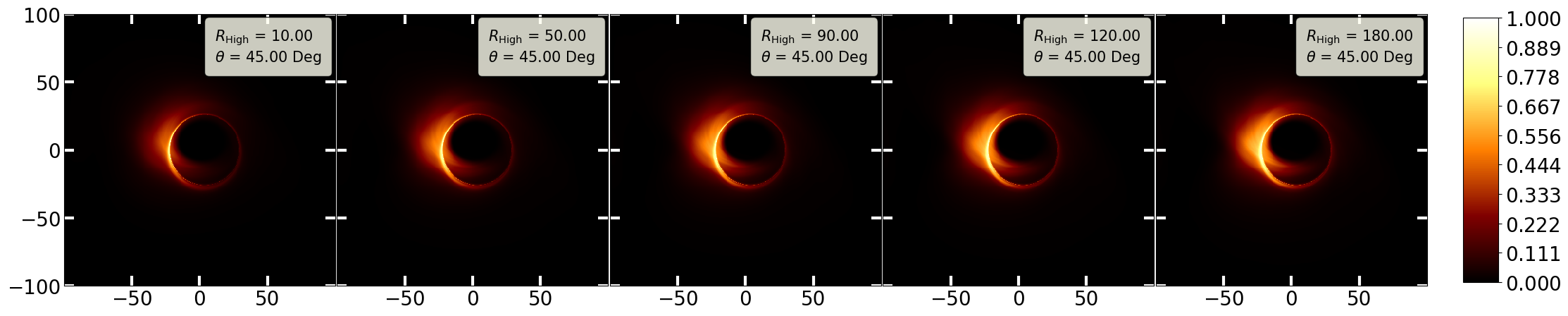}{0.9\textwidth}{(a) $a = +0.5$}}
    \gridline{\fig{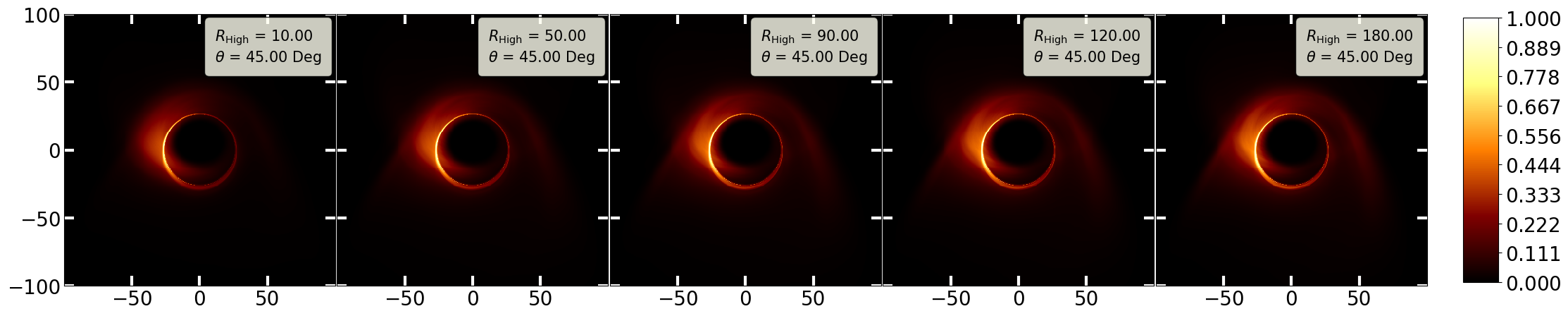}{0.9\textwidth}{(b) $a = 0$}}
    \gridline{\fig{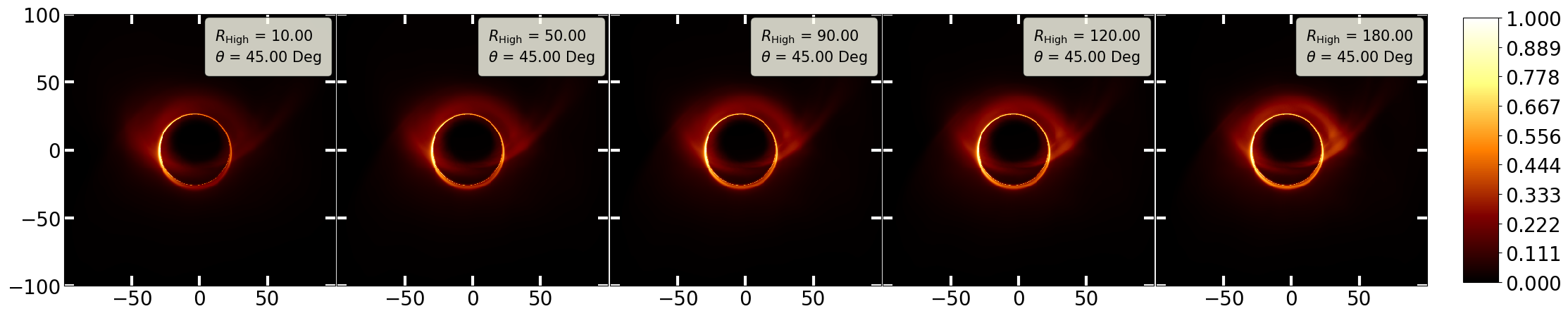}{0.9\textwidth}{(c) $a = -0.5$}}
    \gridline{\fig{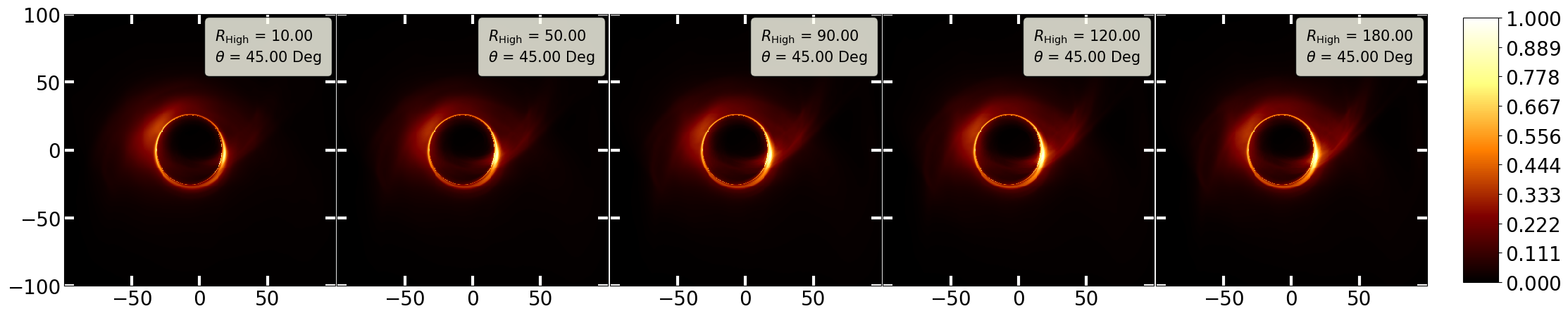}{0.9\textwidth}{(d) $a = -0.94$}}
    \caption{Same as Figure~\ref{fig:lumavg}, but for the representative plots of each black hole spin $a$. We labelled $a$ under the description of each subplot. All images are taken at $\theta = 45$ Deg. \label{fig:lumavgsupp}}
\end{figure*}

\begin{figure*}[htb]
    \centering
    \includegraphics[width=1.0\linewidth]{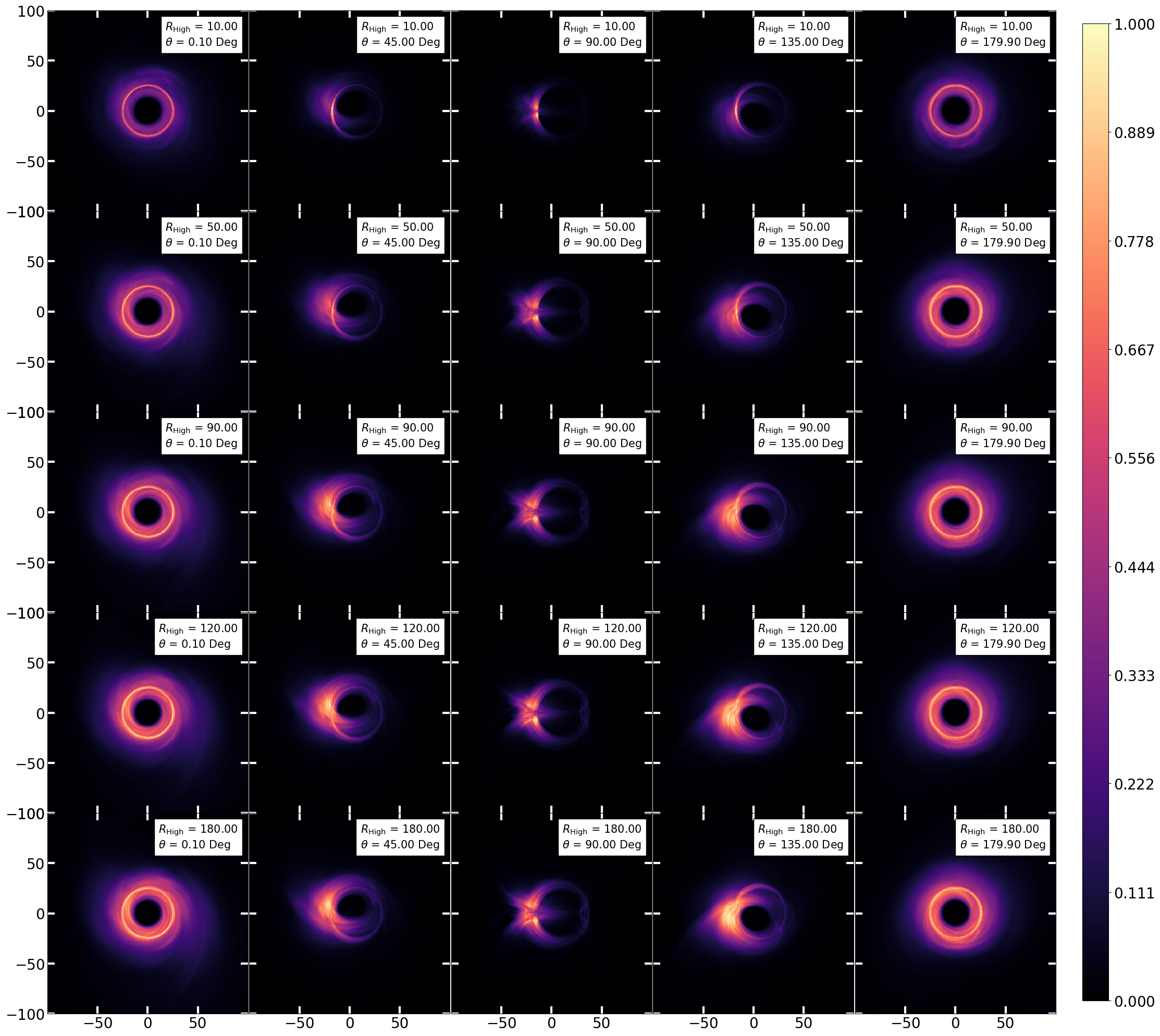}
    \caption{Same as Figure~\ref{fig:lumavg}, but for the pixel-wise, time-domain standard deviation of the luminosity. The standard deviation plots for other spins are included in Figure Set 3. \label{fig:lumstd}}

    \figsetstart
    \figsetnum{3}
    \figsettitle{Pixel-wise, time-domain standard deviation of the luminosity across different $R_{\rm High}$ and $\theta$ (with $R_{\rm Low} = 1$) for $a = +0.5, 0, -0.5,$ and $-0.94$.}
    
    \figsetgrpstart
    \figsetgrpnum{3.1}
    \figsetgrptitle{lumstd+0.5}
    \figsetplot{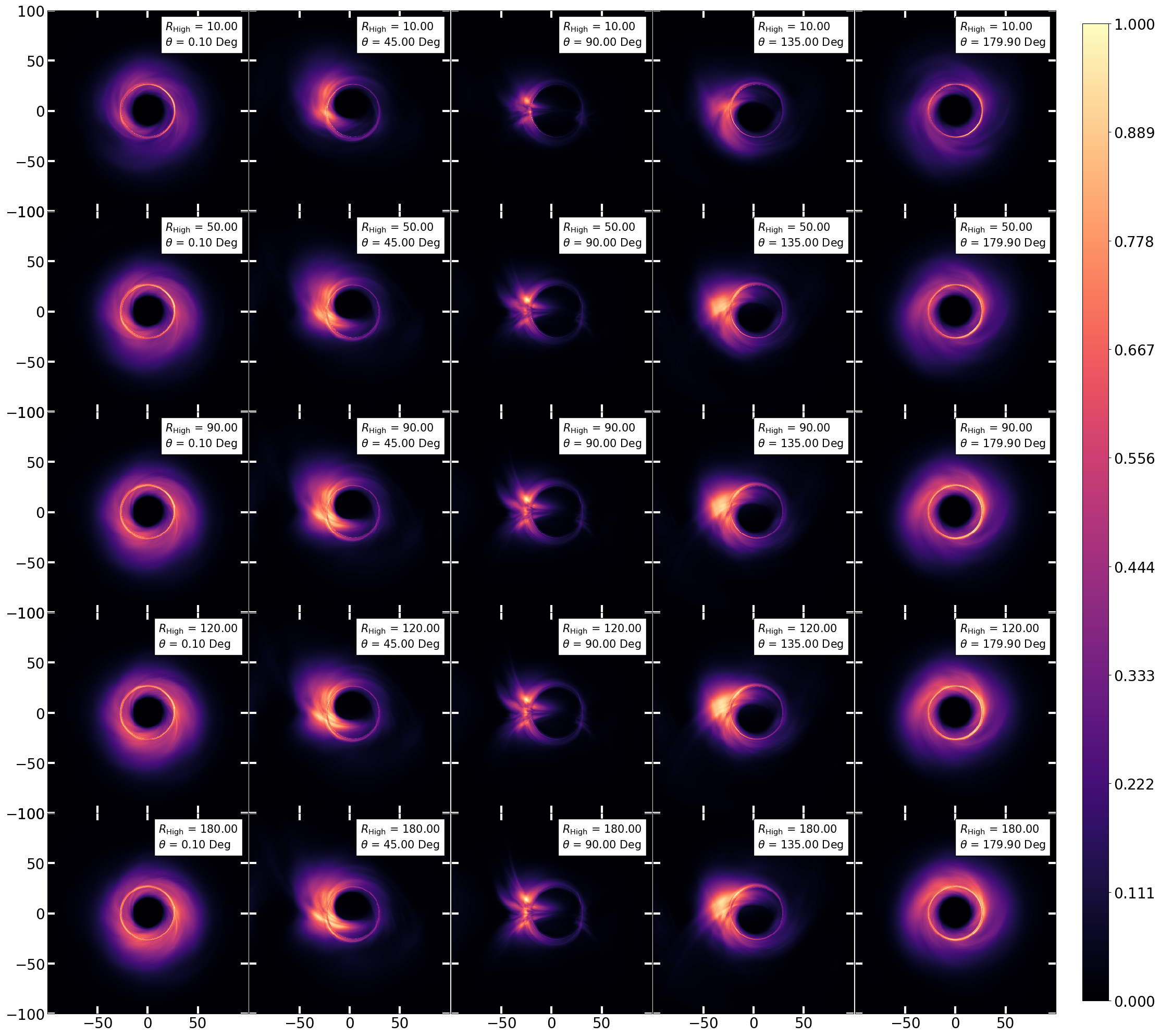}
    \figsetgrpnote{Pixel-wise, time-domain standard deviation of the luminosity across different $R_{\rm High}$ and $\theta$ (with $R_{\rm Low} = 1$) for $a = +0.5$. See Figure~\ref{fig:lumstd} for the details.}
    \figsetgrpend

    \figsetgrpstart
    \figsetgrpnum{3.2}
    \figsetgrptitle{lumstd0}
    \figsetplot{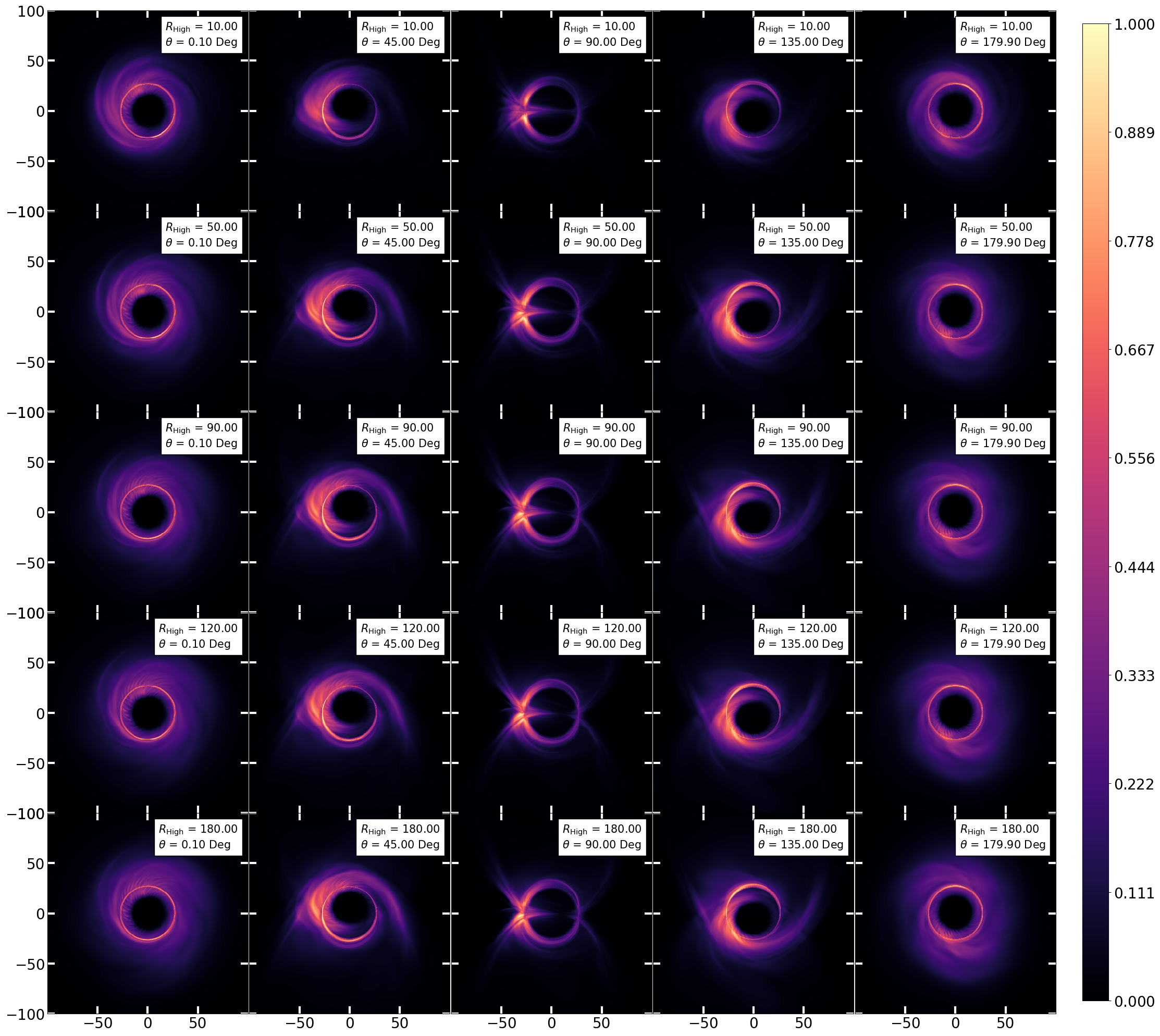}
    \figsetgrpnote{Pixel-wise, time-domain standard deviation of the luminosity across different $R_{\rm High}$ and $\theta$ (with $R_{\rm Low} = 1$) for $a = 0$. See Figure~\ref{fig:lumstd} for the details.}
    \figsetgrpend

    \figsetgrpstart
    \figsetgrpnum{3.3}
    \figsetgrptitle{lumstd-0.5}
    \figsetplot{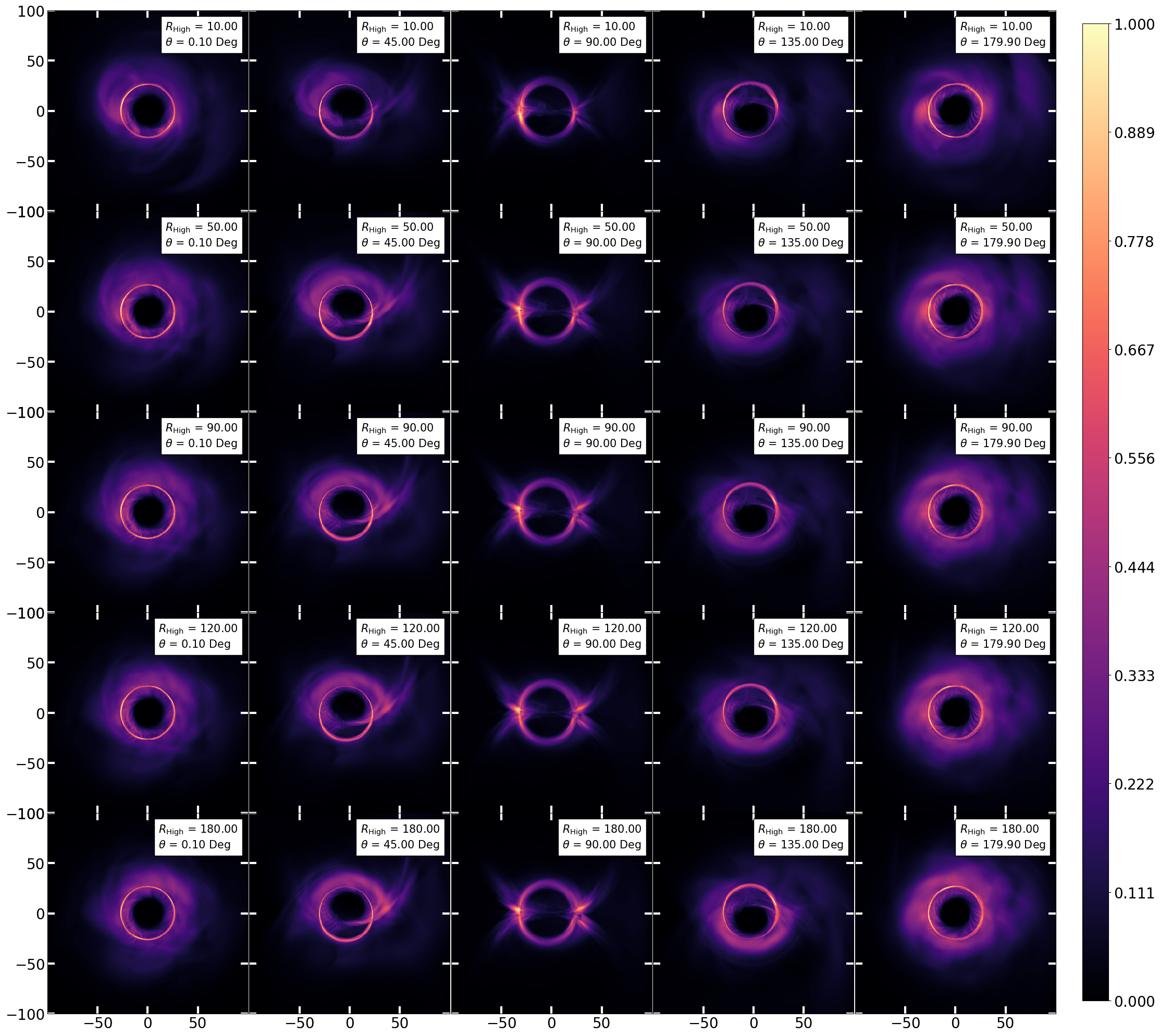}
    \figsetgrpnote{Pixel-wise, time-domain standard deviation of the luminosity across different $R_{\rm High}$ and $\theta$ (with $R_{\rm Low} = 1$) for $a = -0.5$. See Figure~\ref{fig:lumstd} for the details.}
    \figsetgrpend

    \figsetgrpstart
    \figsetgrpnum{3.4}
    \figsetgrptitle{lumstd-0.94}
    \figsetplot{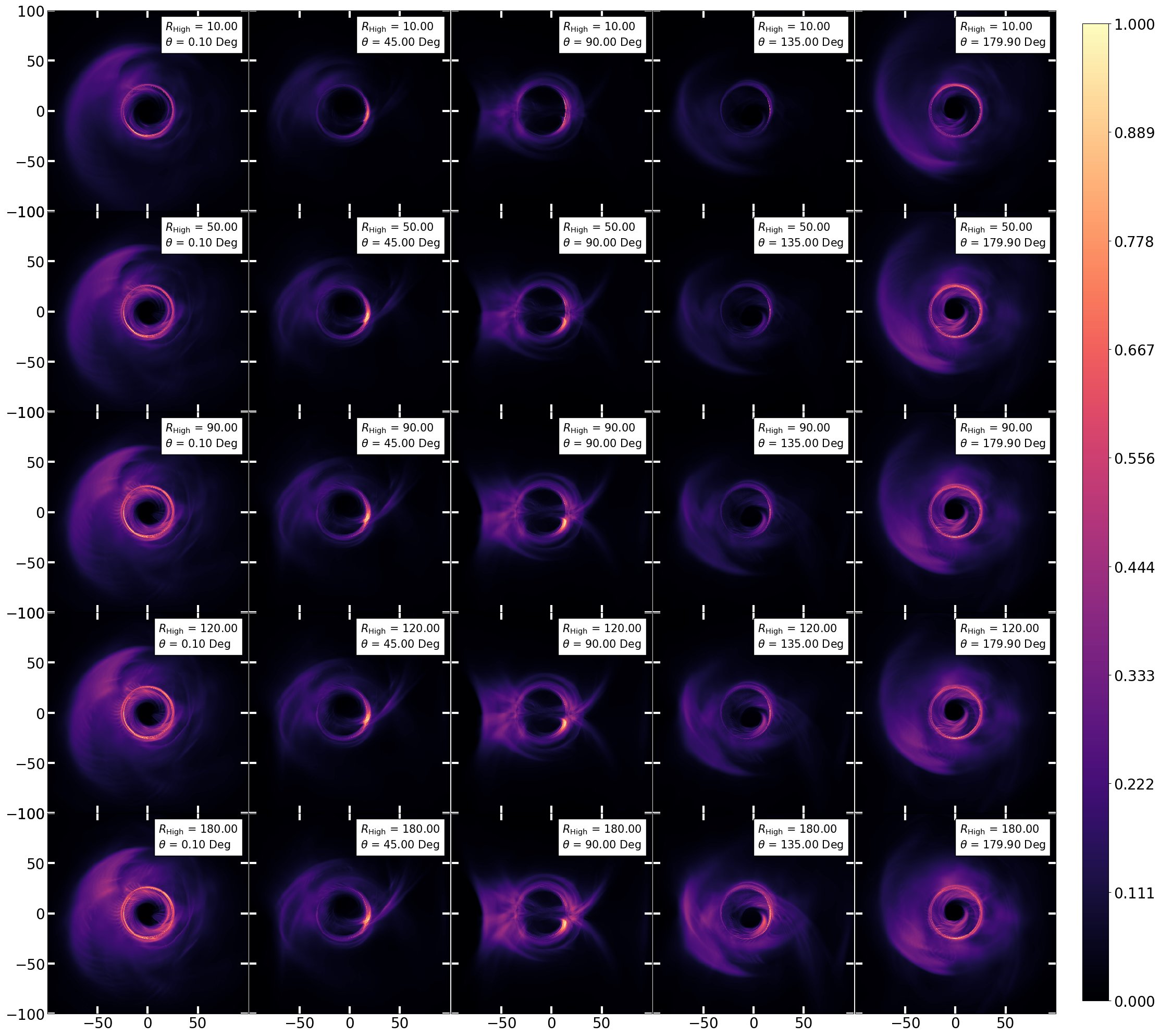}
    \figsetgrpnote{Pixel-wise, time-domain standard deviation of the luminosity across different $R_{\rm High}$ and $\theta$ (with $R_{\rm Low} = 1$) for $a = -0.94$. See Figure~\ref{fig:lumstd} for the details.}
    \figsetgrpend
    
    \figsetend
    
\end{figure*}

\begin{figure*}[htb]
    \centering
    \gridline{\fig{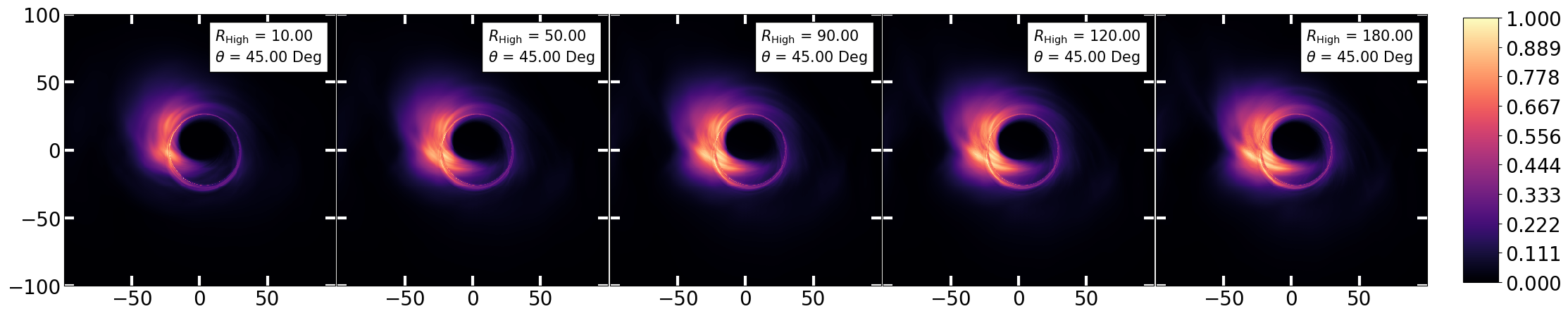}{0.9\textwidth}{(a) $a = +0.5$}}
    \gridline{\fig{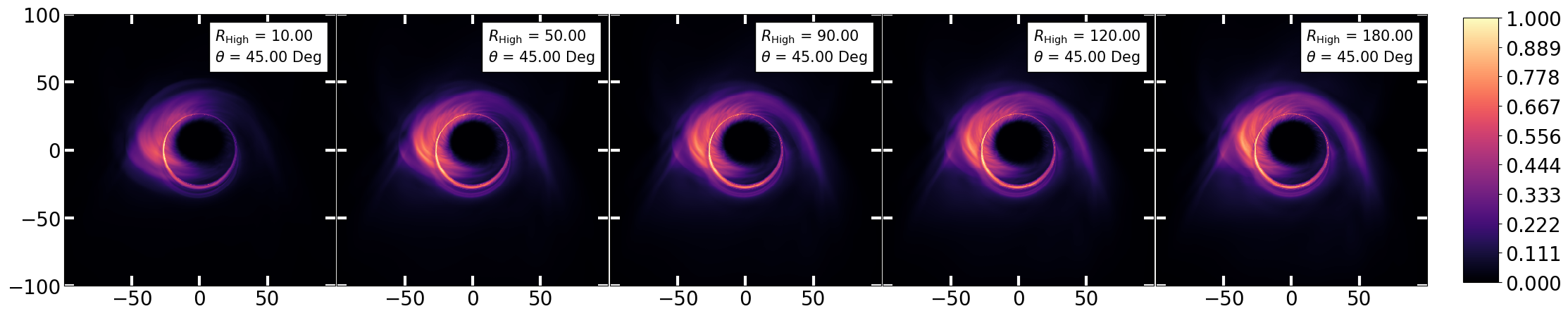}{0.9\textwidth}{(b) $a = 0$}}
    \gridline{\fig{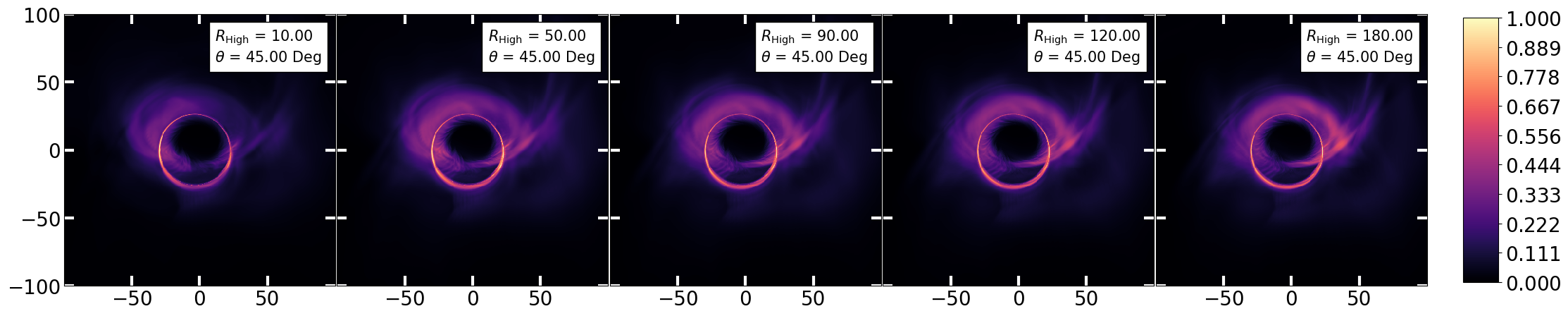}{0.9\textwidth}{(c) $a = -0.5$}}
    \gridline{\fig{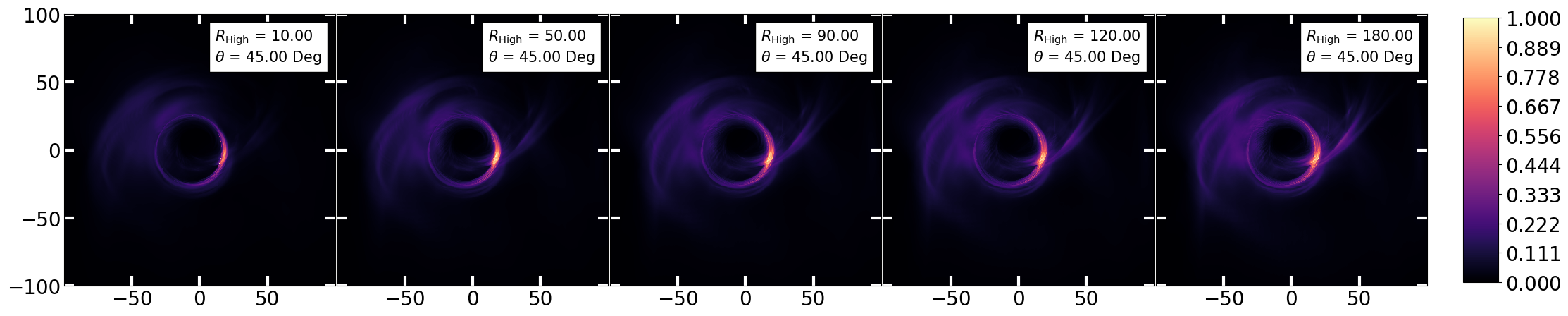}{0.9\textwidth}{(d) $a = -0.94$}}
    \caption{Same as Figure~\ref{fig:lumstd}, but for the representative plots of each black hole spin $a$. We labelled $a$ under the description of each subplot. All images are taken at $\theta = 45$ Deg. \label{fig:lumstdsupp}}
\end{figure*}

%%%%%%%%%%%%%%%%%%%%%%%%%%%%%%%%%%%%%%%%%%%%%%%%%%%%%%%%%%%%%%%%%%%%%%%%%%%%%%%%%%%%%%%%%%%%%%%%%%%%%%%%%%%%%%%%%%%%%

\subsection{Model Parameters} \label{subsec:models} 

\begin{deluxetable}{cc}
\caption{List of parameters of the GRMHD models and the GRRT parameter surveys we considered. \label{tab:params}}
\tablewidth{0pt}
\tablehead{
\colhead{Parameters} & \colhead{Range}
}
\startdata
Black-hole Spin $a$ & ($0.94$, $0.5$, $0$, $-0.5$, $-0.94$) \\
R$_{\rm High}$ & ($10$, $50$, $90$, $120$, $180$) \\
Inclination $\theta$ (Deg) & ($0.1$, $45$, $90$, $135$, $179.9$) \\
$R_{\rm Low}$ & ($1$ -- $\text{MIN}[60, R_{\rm High}]$) \\
$M_{\rm Unit}$ & ($10^{17}$ -- $10^{21}$) \\
$\tau$ & ($\textrm{29,465}$ -- $\textrm{29,995}$)\,$GMc^{-2}$ 
\enddata
\end{deluxetable}

To investigate the effect of varying $R_{\text{Low}}$ to $M_{\Delta T}$, we perform GRRT parameter surveys on the open science grid \citep{osg07, osg09, https://doi.org/10.21231/906p-4d78}, which is a set of pools of shared computing and data capacity for distributed high-throughput computing, supported by the National Science Foundation and the Department of Energy. Table~\ref{tab:params} shows the parameters we considered in this study, which include the spin of the black of $a$, $R_{\rm High}$, and the inclination angle of the observer $\theta$. In particular, we consider the time interval $\tau = (\textrm{29,465}$ -- $\textrm{29,995})\,GMc^{-2}$ of snapshots, which covers a duration of $530\,GMc^{-2} \sim 3\,\mathrm{hours}$, to compute $M_{\Delta T}$. This interval represents the last sets of snapshots of the simulations, in which inflow equilibrium near the horizon, where the $230$\,GHz emissions are dominated, shall be established. In general, GRMHD simulations of accreting systems are scale-invariant, so there is a free parameter $M_{\rm Unit}$ that scales the gas density of the system. $M_{\rm Unit}$ converts from the code unit to the cgs unit. We consider $10$ sets of $M_{\rm Unit}$ ranging from $10^{17}$ to $10^{21}$ (in the log scale), which spans the optically thin to optically thick regime. Additionally, we consider $10$ sets of $R_{\rm Low}$ for a given $R_{\rm High}$, ranging from $R_{\rm Low} = 1$ to $R_{\rm Low} = \text{MIN}(60, R_{\rm High})$. We set an upper limit for $R_{\rm Low}$ because the black-hole luminosity drops quickly at large $R_{\rm Low}$, where the highly magnetized electron becomes cold. Additionally, as we will show in the later section, an increase in $R_{\rm Low}$ accompany an increase in  Thus, $R_{\rm Low}$ should not increase further, or the assumption of Sgr~A* being optically thin will be broken. Altogether, given a certain set of $a$, $R_{\rm High}$, $\theta$, and a particular GRMHD snapshot, we would have $10 \times 10 = 100$ of GRRT images spanning different $R_{\rm Low}$ and $M_{\rm Unit}$. 

After having the black-hole flux at $230$\,GHz for a particular set of $a$, $R_{\rm High}$, $\theta$, and in our considered time interval $\tau$, we can then compute the time-domain average, variance, and hence, the variability for the $230$\,GHz flux. We could then infer how $R_{\rm Low}$ should vary against $M_{\rm Unit}$ for black-hole models that satisfy the constraint of the Sgr~A* $230$\,GHz mean flux, which is 2.4\,Jy \citep{2022ApJ...930L..19W}. In addition, we would analyze GRRT images satisfying the flux constraint, and from the combined analysis, we can hopefully answer the following:

\begin{enumerate}
    \item Can varying $R_{\rm Low}$ reduce the $230$\,GHz variability?
    \item Why are the current GRMHD models too variable at $230$\,GHz compared to observations?
\end{enumerate}

Note, however, that we assumed the vertical axis of the camera aligns with the vertical axis of the spinning black hole. 

%%%%%%%%%%%%%%%%%%%%%%%%%%%%%%%%%%%%%%%%%%%%%%%%%%%%%%%%%%%%%%%%%%%%%%%%%%%%%%%%%%%%%%%%%%%%%%%%%%%%%%%%%%%%%%%%%%%%%

\section{Results and Discussion} \label{sec:results}

This section presents and discusses the results of our GRRT parameter surveys, the dependencies of $M_{\Delta T}$ on $R_{\rm Low}$ and $M_{\rm Unit}$, and the origin of the high variability for the $R_{\rm Low} = 1$ images. Note that we refer to prograding (retrograding) black holes as black holes with positive spin $a > 0$ (negative spin $a < 0$) and that non-rotating black holes are referred to as non-spinning $a = 0$ black holes.

%%%%%%%%%%%%%%%%%%%%%%%%%%%%%%%%%%%%%%%%%%%%%%%%%%%%%%%%%%%%%%%%%%%%%%%%%%%%%%%%%%%%%%%%%%%%%%%%%%%%%%%%%%%%%%%%%%%%%

\subsection{Parameter Surveys} \label{subsec:param}

\textbf{Could $M_{\Delta T}$ be lowered with varying $R_{\rm Low}$?} Here, we include results for the GRMHD model of $a = +0.94$ in Figure~\ref{fig:m3contour} as a reference. We also include some representative plots for other black-hole spins in Figure~\ref{fig:m3contoursupp}. At the same time, the remaining full results are shown in Figure Set 1. The light-blue markers indicate the models we obtained from GRRT. The black-dotted lines in the contour plots of Figure~\ref{fig:m3contour} represent models consistent with the Sgr~A* $230$\,GHz flux. The green lines indicate regions of the parameter spaces where $M_{\Delta T} \leq 0.1$. This is also the upper limit of the historical distribution of $M_{\Delta T}$ for Sgr~A*. Depending on the black-hole spins, the parameter space where $M_{\Delta T} \leq 0.1$ could be consistent with the Sgr~A* flux constraint. For instance, black holes with $a = +0.94$ and $a = +0.5$ (for $\theta = 90$ Deg) show a limited sequence of Sgr~A* models that intersect with the region bounded by the green line ($M_{\Delta T} \leq 0.1$), which are obtained by increasing $R_{\rm Low}$ beyond $1$. Non-spinning black-hole, except for $R_{\rm High} = 10$, contains sequences of $M_{\Delta T} \leq 0.1$ that is consistent with the Sgr~A* flux. We also note that all $R_{\rm High} = 10$ Sgr~A* models with $a = +0.94$ and $+0.5$ could not attain $M_{\Delta T} \leq 0.1$ irrespective of $R_{\rm Low}$ assumed. 

However, black holes with $a < 0$ differ from those with $a \geq 0$. Except for $R_{\rm High} = 120, 180$ viewed at $\theta = 0.1\,\mathrm{Deg}$, black-hole with $a = -0.5$ that have $M_{\Delta T} \leq 0.1$ are inconsistent with the Sgr~A* flux constraint. In fact, we find that no parameter spaces with $M_{\Delta T} \leq 0.1$ that are consistent with the observed $M_{\Delta T}$ for Sgr~A* with $a = -0.94$. This suggests a substantial reduction in $M_{\Delta T}$ with varying $R_{\rm Low}$ is model-dependent and sensitive to the black-hole parameters. This also suggests that the effects of varying $R_{\rm Low}$ to the $230$\,GHz image are different between $a \geq 0$ and $a < 0$ black holes, and we will compare and explain the differences in more detail in our next papers. 

In the contour sub-plots, we find an almost log-linear relationship between $R_{\rm Low}$ and $M_{\rm Unit}$ for suites of Sgr~A* models for all of the $\theta$ and $R_{\rm High}$ parameter values we considered. This could be interpreted qualitatively: as $R_{\rm Low}$ increases, the thermal energy partitioned to the electrons at low $\beta$ becomes smaller, making the electrons cooler. Note that the major contribution to the $230$\,GHz flux is synchrotron radiation, for which its emissivity is an increasing function of $T_{e}$ and $\rho$. Given the $2.4\,\textrm{Jy}$ constraint, the fluid density needs to increase to compensate for a reduction in $T_{e}$. This explains the observed trend for $M_{\rm Unit}$. We note a similar pattern between $R_{\rm Low}$ and $M_{\rm Unit}$ for other black-hole spins.

%%%%%%%%%%%%%%%%%%%%%%%%%%%%%%%%%%%%%%%%%%%%%%%%%%%%%%%%%%%%%%%%%%%%%%%%%%%%%%%%%%%%%%%%%%%%%%%%%%%%%%%%%%%%%%%%%%%%%

\subsection{Parameter Dependence} \label{subsec:depdend}

\textbf{How would $M_{\Delta T}$ vary as $R_{\rm Low}$ for suites of Sgr~A* models?} We extract the values of $M_{\Delta T}$ and $R_{\rm Low}$ along the black-dotted lines in Figure~\ref{fig:m3contour} and present them in Figure~\ref{fig:minm3}. To extract the values of $M_{\Delta T}$, we perform Bivariate spline approximations with default parameters provided by the package \texttt{Scipy}, and we did it for all the spins we considered. In sum, we find three different patterns for the $M_{\Delta T}$ versus $R_{\rm Low}$ curves: \emph{i}) $M_{\Delta T}$ increases as $R_{\rm Low}$, \emph{ii}) $M_{\Delta T}$ decreases as $R_{\rm Low}$, and \emph{iii}) $M_{\Delta T}$ first decreases to a local minimum and then increases. We observe that except for the case of $R_{\rm Low} = 10$, the shape of the $M_{\Delta T}$ versus $R_{\rm Low}$ curves is less sensitive to the variations of $R_{\rm Low}$, but is more sensitive to the changes in $\theta$. 

Almost every $a \geq 0$ black-hole model, with $R_{\rm High} > 10$, shows pattern \emph{iii} as $R_{\rm Low}$ first increases beyond $1$. Whether or not $M_{\Delta T}$ reduces for $a < 0$ black-hole models depends on $a$ and $\theta$. Also, the $M_{\Delta T}$ versus $R_{\rm Low}$ curves for $a < 0$ black holes differ from those of $a > 0$ black holes. For instance, none of the $a < 0$ black-hole curves shows strong dips in the $M_{\Delta T}$ versus $R_{\rm Low}$ relations, and they vary more slowly and smoothly than those of the $a > 0$ black holes. This also proves that the GRRT images of changing $R_{\rm Low}$ for the $a < 0$ black holes fundamentally differ from those for the $a > 0$ black holes. 

We also find that some suites of Sgr~A* models with $R_{\rm Low} > 10$ show a reduction in $M_{\Delta T}$ as $R_{\rm Low}$ increases. For instance, it is possible to achieve $M_{\Delta T} \lesssim 0.1$ for almost every model with spin $a = +0.94$ and $a = 0$. This result contrasts those presented in \citet{2022ApJ...930L..16E} and is discovered by considering a more comprehensive range of parameter spaces. Thus, we successfully show that it is possible to reduce $M_{\Delta T}$ by varying $R_{\rm Low}$. We note, however, that $M_{\Delta T}$ remains larger than $0.1$ for almost every model for the remaining spins. Thus, whether or not a substantial reduction in $M_{\Delta T}$ occurs when $R_{\rm Low}$ is varied along suites of Sgr~A* models highly depends on the underlying model parameters. 

%%%%%%%%%%%%%%%%%%%%%%%%%%%%%%%%%%%%%%%%%%%%%%%%%%%%%%%%%%%%%%%%%%%%%%%%%%%%%%%%%%%%%%%%%%%%%%%%%%%%%%%%%%%%%%%%%%%%%

\subsection{Large $M_{\Delta T}$ Uncovered - Image Domain} \label{subsec:large}

\textbf{What are the origins of large $M_{\Delta T}$ for the GRRT images of Sgr~A* at $R_{\rm Low} = 1$} (the value assumed in \citealt{2022ApJ...930L..16E})? To understand this, we computed the time domain, pixel-wise \emph{i}) averaged and \emph{ii}) standard deviation of image luminosity across all $a$, and $\theta$. We show the time-averaged, pixel-wise luminosity in Figure~\ref{fig:lumavg} for the $a = +0.94$ GRMHD models. We also include representative plots for other black-hole spin in Figure~\ref{fig:lumavgsupp}. Results for the remaining $a$ are shown in Figure Set 2. The disk is optically thin. Hence, for $\theta = 0.1, 179.9\,\mathrm{Deg}$, the GRRT images show a thin photon ring. As $\theta$ progresses towards $90$ Deg, most of the emission is concentrated to the left of the black-hole shadow due to the Doppler beaming effect. We find that for all the $R_{\rm Low} = 1$ images, the emissions are contributed from a thin photon ring. This is true even for other spins $a$. 

Having all emissions contributed from a small area indicates that a slight variation in the photon ring luminosity would induce large variability in the $230$\,GHz flux. Indeed, we show the time-domain, pixel-wise standard deviation of the $230$\,GHz luminosity for the $a = +0.94$ GRMHD models in Figure~\ref{fig:lumstd}. We also include representative plots for other black-hole spin in Figure~\ref{fig:lumstdsupp}. Results for the remaining $a$ are shown in Figure Set 3. We find that the photon rings contribute most to the image variability. However, some minor contributions from the accretion flow could also be seen. This is true also for the $a = +0.5$ and $0$ models.

On the other hand, the situation for $a < 0$ black holes is different from that of $a \geq 0$ black holes. In addition to the photon-ring variability, the contribution to the flux variability from the accretion flow is more significant for the $a < 0$ black-hole models, and their contribution is the largest for the $a = -0.94$ model. This could be due to the misalignment between the black-hole and torus angular momentum, creating highly irregular accretion flow dynamics.

We note that the image asymmetry due to the Doppler beaming effect is reduced as the black hole is spinning more negatively. Such an effect is first mentioned in \citet{medeiros2022brightness}, where they consider image asymmetry up until $a = 0$. Here, we briefly extend their work to consider also $a < 0$ black holes. The image becomes more symmetric for $a = -0.5$ and $a = -0.94$. This is probably due to the stronger retrograde motion black hole, which drags the rotating plasma in the opposite direction to the plasma angular velocity, reducing the angular velocity differences between the plasma on the left and right-hand side of the event horizon and thus weakening the Doppler beaming effect.

%%%%%%%%%%%%%%%%%%%%%%%%%%%%%%%%%%%%%%%%%%%%%%%%%%%%%%%%%%%%%%%%%%%%%%%%%%%%%%%%%%%%%%%%%%%%%%%%%%%%%%%%%%%%%%%%%%%%%

\subsection{Matching Observed $M_{\Delta T}$ for Sgr~A*} \label{subsec:smalldelta}

\begin{figure}[htb!]
    \centering
    \includegraphics[width=1.0\linewidth]{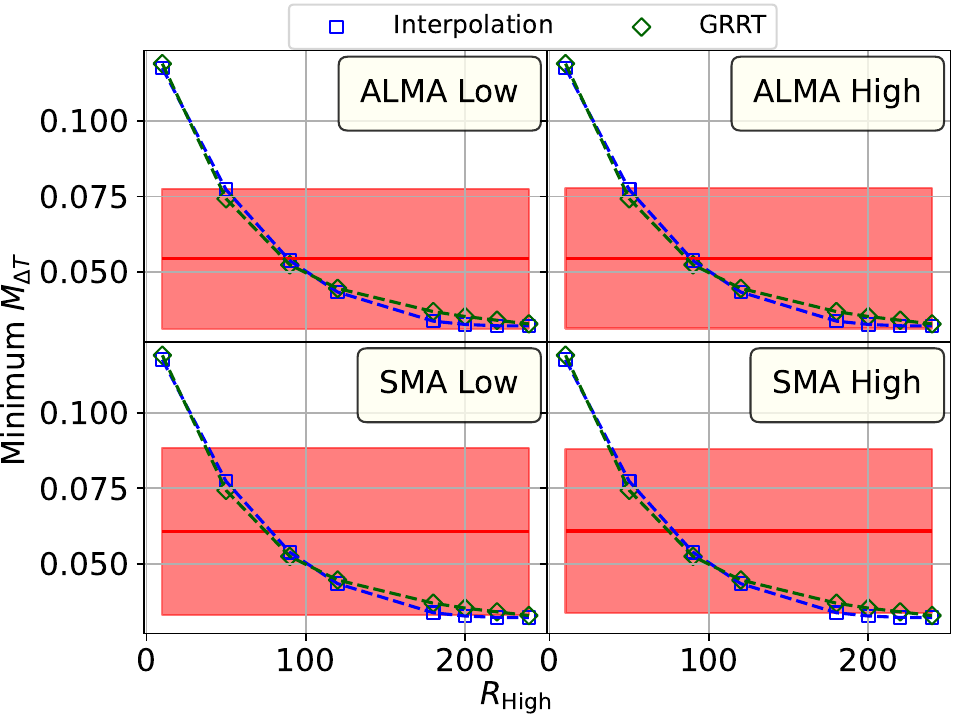}
    \caption{The variation of the minimum of $M_{\Delta T}$, obtained through varying $R_{\rm Low}$ from $1$ to $60$, against a given $R_{\rm High}$. They are shown as scatter points. This is for the particular black-hole parameter with $a = + 0.94$ and $\theta = 0.1\,\mathrm{Deg}$, which shows exceptionally small $M_{\Delta T}$. In each subplot, the upper right box shows the telescope data and the observational band filter that we use to compute the $3\,\mathrm{hours}$ variability. The solid red horizontal lines and the red shaded represent the observed mean and standard deviation of $M_{\Delta T}$, respectively. We use Bivariate spline approximations to obtain the blue-square scatter points while using direct GRRT to compute the green-diamond scatter points. \label{fig:smallm3}}
\end{figure}

\textbf{Are there models consistent with the observed $3\,\mathrm{hours}$ variability of Sgr~A*?} In Section~\ref{subsec:param}. we find that it is possible to lower $M_{\Delta T}$ if we vary $R_{\rm Low}$ for certain sets of model parameters. Additionally, we find from Figure~\ref{fig:minm3} that the minimum attainable $M_{\Delta T}$ across different $R_{\rm Low}$ for spin $a = +0.94$ and $\theta = 0.1\,\mathrm{Deg}$ decreases as $R_{\rm High}$. This motivates us to perform extra parameter searches for the model parameters $a = +0.94$ and $\theta = 0.1\,\mathrm{Deg}$ with $R_{\rm High}$ beyond $180$ and extract the corresponding minimum $R_{\rm Low}$. We aim to find whether $M_{\Delta T}$ could be further lowered and if they are consistent with the observed $3\,\mathrm{hours}$ variability of Sgr~A*.

In particular, we consider $R_{\rm High} = 200, 220$, and $240$. We performed GRRT with varying $R_{\rm Low}$ given these $R_{\rm High}$ and to look for models that satisfy the $2.4$\,Jy constraints. We then extract the corresponding minimum of $M_{\Delta T}$. We use two methods to achieve this. The first is to directly look for the minimum of $M_{\Delta T}$ along the curves generated by the Bivariate spline approximations (c.f. Figure~\ref{fig:minm3}). The second is to look for the value of $R_{\rm Low}$ (and $M_{\rm Unit}$) that produces such a $M_{\Delta T}$, and we further perform GRRT to compute the corresponding $M_{\Delta T}$. In addition, we extract ALMA and SMA observations of Sgr~A* at $230$\,GHz and compute the mean and standard deviation of the observed $M_{\Delta T}$. To do this, we perform windowed sampling with a window size of $3\,\mathrm{hours}$, and and that the starting point of this window moves forward by $0.5\,\mathrm{hours}$ for each sampling. We also note that the ALMA and SMA observations are performed with the High and Low bands. We compute the $M_{\Delta T}$ separately. 

These results are shown in Figure~\ref{fig:smallm3}. We find that all $R_{\rm High} = 10$ models are inconsistent with the $M_{\Delta T}$ constraint even if we vary $R_{\rm Low}$. $R_{\rm High} \geq 50$ models could attain $M_{\Delta T}$ consistent with the observations. As $R_{\rm High}$ increases beyond $180$, the minimum of $M_{\Delta T}$ could be as low as $\sim 0.03$. Also, we find that the minimum of $M_{\Delta T}$ decreases for a larger $R_{\rm High}$, but the reduction rate also decreases. Indeed, the minimum of $M_{\Delta T}$ for $R_{\rm High} = 240$ is close to that of $R_{\rm High} = 180$. This might suggest a threshold value of $M_{\Delta T}$, where $M_{\Delta T}$ could not be lowered than such value even if $R_{\rm Low}$ is considered. More comprehensive statistical studies will help reveal whether such value exists.

%%%%%%%%%%%%%%%%%%%%%%%%%%%%%%%%%%%%%%%%%%%%%%%%%%%%%%%%%%%%%%%%%%%%%%%%%%%%%%%%%%%%%%%%%%%%%%%%%%%%%%%%%%%%%%%%%%%%%

\section{Conclusion} \label{sec:conclu}

In this paper, we performed a comprehensive set of GRRT parameter surveys using the Open Science Grid to study the reason for the unexpectedly high variability $M_{\Delta T}$ of Sgr~A* predicted by theoretical models in \citet{2022ApJ...930L..16E}. Our parameter surveys vary the $R_{\rm Low}$ parameter in the electron temperature prescription function used by \citet{2022ApJ...930L..16E}, which spans the optically thin to the optically thick regimes, and covers $R_{\rm Low}$ from $1$ to $60$. Our results contrast to those in \citet{2022ApJ...930L..16E}. We find that it is possible to reduce $M_{\Delta T}$ to a value lower than $0.1$, the threshold values of the historical distribution of $M_{\Delta T}$ for Sgr~A*, by varying $R_{\rm Low}$, for specific sets of black-hole spin $a$, inclination angle $\theta$, and $R_{\rm High}$. 

We find that models with $a = +0.94$, $\theta = 0.1\,\mathrm{Deg}$, and $R_{\rm High} \geq 50$ could produce a low-level of variability consistent with the observed $M_{\Delta T}$ of Sgr~A*. We also note that the minimum of $M_{\Delta T}$ reduces for larger $R_{\rm High}$ for this particular set of $a$ and $\theta$, and $M_{\Delta T} \sim 0.03$ for $R_{\rm High} \geq 180$. Although this might be inconsistent with the statistical average of $M_{\Delta T}$, we note that the minimum of $M_{\Delta T}$ across the historical observational distribution could be as low as $0.03$. Thus, such a result sheds light on whether $R_{\rm Low}$ could be an important parameter to consider if one intends to perform GRRT on suites of Sgr~A* simulation libraries. 

We also analyze the GRRT images for the $R_{\rm Low} = 1$ models, which are also the values assumed in \citet{2022ApJ...930L..16E}, across different $a$ and $\theta$. By computing the time domain, pixel-wise, averaged, and standard deviation of the image luminosity, we find that all $R_{\rm Low} = 1$ models have their luminosity contributions to be dominated by a thin photon ring. We also find that the thin photon ring contributes most to the observed variability for all $a \geq 0$ models. For $a < 0$ black-hole models, secondary contributions of variability from the accretion flow are more visible due to the irregular flow dynamics induced by misalignment between the black-hole and torus angular momentum. Our findings on the origin of high $M_{\Delta T}$ suggest that a more serious study on the electron temperature used to model RIAF should be adopted. These results, together with the findings that the behaviour of $M_{\Delta T}$ versus $R_{\rm Low}$ relations are different between the $a < 0$ and $a \geq 0$ black holes, also lay the foundations for our successive papers, where we would analyze why a variation of $R_{\rm Low}$ could reduce/increase $M_{\Delta T}$ for different black-hole parameters. 

Note that there are still several caveats in this preliminary study. We assumed only $R_{\rm Low} \leq 60$ in our study. This might raise concerns about the lack of details on the behaviour of $M_{\Delta T}$ against $R_{\rm Low}$ at a larger $R_{\rm Low}$. However, we have already shown that $M_{\Delta T}$ would decrease in the range of parameter spaces we are interested in. Thus, there seems to be no immediate need to consider higher $R_{\rm Low}$. Also, we show an almost log-linear relationship between $R_{\rm Low}$ and $M_{\rm Unit}$ for suites of Sgr~A* models. Thus, a further increase in $R_{\rm Low}$ would make the accretion disk more optically thick. This would violate our assumption of Sgr~A* being a low-luminosity, geometrically thick, and optically thin accretion flow. 

Computing the effects of $R_{\rm Low}$ on $M_{\Delta T}$ for one particular given time interval $\tau = (\textrm{29,465}$ -- $\textrm{29,995})\,GMc^{-2}$ only might induce concerns about whether our results are statistically significant. Also, coarse parameter spaces that we considered in our work might pose worries about whether the minimum of $M_{\Delta T}$ is due to interpolation errors. However, we show in Figure~\ref{fig:smallm3} that the minimum of $M_{\Delta T}$ computed by the Bivariate spline approximations and direct GRRT are close. Nonetheless, these limitations are due to the limited computational storage and time constraints. We plan to perform more comprehensive studies to understand better the role of $R_{\rm Low}$ in a more serious statistical framework on the increase/decrease of $M_{\Delta T}$.

%%%%%%%%%%%%%%%%%%%%%%%%%%%%%%%%%%%%%%%%%%%%%%%%%%%%%%%%%%%%%%%%%%%%%%%%%%%%%%%%%%%%%%%%%%%%%%%%%%%%%%%%%%%%%%%%%%%%%

\begin{acknowledgments}
We thank the anonymous referee and Dr. David Hughes for providing valuable comments that improve the manuscript. Ho Sang (Leon) Chan acknowledges support from the Croucher Scholarship for Doctoral Studies by the Croucher Foundation. C.-k.~C. acknowledge NSF grant OISE 17-43747 support.
\end{acknowledgments}

%%%%%%%%%%%%%%%%%%%%%%%%%%%%%%%%%%%%%%%%%%%%%%%%%%%%%%%%%%%%%%%%%%%%%%%%%%%%%%%%%%%%%%%%%%%%%%%%%%%%%%%%%%%%%%%%%%%%%

\bibliography{main}{}
\bibliographystyle{aasjournal}

%%%%%%%%%%%%%%%%%%%%%%%%%%%%%%%%%%%%%%%%%%%%%%%%%%%%%%%%%%%%%%%%%%%%%%%%%%%%%%%%%%%%%%%%%%%%%%%%%%%%%%%%%%%%%%%%%%%%%

\end{document}